\documentclass[12pt,preprint]{aastex}

\usepackage{epsfig}

\def\msol{\hbox{$M_\odot$}}

\def\e10{\eta_{10}}

\def\etal{{\it et al.\ }}
\def\iso#1#2{\mbox{${}^{#2}{\rm #1}$}}
\newcommand\he[1]{\iso{He}{#1}}
\newcommand\li[1]{\iso{Li}{#1}}

\def\li#1{\iso{Li}{#1}}
\def\b1#1{\iso{B}{1#1}}
\def\mg2#1{\iso{Mg}{2#1}}
\def\msun{\mbox{M$_\odot$}}

\def\beq{\begin{equation}}
\def\eeq{\end{equation}}
\def\beqar{\begin{eqnarray}}
\def\eeqar{\end{eqnarray}}
\def\simlt{\lower.5ex\hbox{$\; \buildrel < \over \sim \;$}}
\def\simgt{\lower.5ex\hbox{$\; \buildrel > \over \sim \;$}}
\def\simpropto{\lower.2ex\hbox{$\; \buildrel \propto \over \sim \;$}}

\begin{document}

\title{Cosmic Chemical Evolution with an Early  Population of Intermediate Mass Stars}
 
\author{Elisabeth Vangioni}
\affil{Institut d'Astrophysique de Paris, UMR 7095 CNRS, University Pierre et Marie Curie, 98 bis Boulevard Arago,
Paris 75014, France}

\author{Joseph Silk}
\affil{Department of Physics, University of Oxford, Keble Road, Oxford OX1 3RH,\\
and Institut d'Astrophysique, 98 bis Boulevard Arago,
Paris 75014, France}

\author{Keith~A.~Olive}
\affil{ Theoretical Physics
Institute, School of Physics and Astronomy, \\
University of Minnesota, Minneapolis, MN 55455 USA}

 \and \author{Brian D. Fields}
\affil{Departments of Astronomy and of Physics, University of Illinois,
Urbana, IL 61801, USA}

\begin{abstract}

\vskip-6in
\begin{flushright}
UMN-TH-2923/10 \\
TPI-MINN-10/30 \\
October 2010
\end{flushright}
\vskip+5.0in

We explore the consequences of an early population of intermediate mass stars
in the 2 -- 8 M$_\odot$
range on cosmic chemical evolution.
We discuss the implications of this population as it pertains to
several cosmological and astrophysical observables. 
For example, some very metal-poor 
galactic stars show large enhancements of carbon,
typical of the C-rich ejecta of low-mass stars but not of supernovae;
moreover, halo star carbon and oxygen abundances show wide
scatter, which imply a wide range of
star-formation and nucleosynthetic histories contributed
to the first generations of stars.
Also, recent analyses of the \he4 abundance in metal-poor 
extragalactic H II regions suggest an elevated abundance
$Y_p \simeq 0.256$ by mass, higher than the 
predicted result from big bang nucleosynthesis assuming the 
baryon density determined by WMAP, $Y_p = 0.249$.
Although there are large uncertainties in the observational
determination of \he4, this offset may suggest a prompt initial enrichment
of \he4 in early metal-poor structures.
We also discuss the effect of intermediate mass stars on
global cosmic evolution, the reionization of the Universe, 
the density of white dwarfs, 
as well as SNII and  SNIa rates at high redshift.
We also comment on the early astration of D and \li7.
We conclude that if intermediate mass stars are to be associated with 
Population III stars, their relevance is limited (primarily from observed
abundance patterns) to low mass structures involving a limited fraction of the 
total baryon content of the Universe.

\clearpage

\end{abstract}


\section{Introduction}

One of the outstanding questions concerning the first epoch of star formation
is the stellar mass distribution of Population III stars.  Most often, Pop III stars
are assumed to be very massive (40 - 100 M$_\odot$) or supermassive
($>$ 100 M$_\odot$). Tracing chemical abundance patterns through
cosmic chemical evolution allows one to test  theories employing massive Pop III
stars.  For example, the abundance patterns seen in extremely iron-poor stars
do not support the hypothesis that the first stars had masses between $ 140 - 250$ M$_\odot$
and end as pair-instability supernovae \citep{venk,daigne1}. 
Yet other observations, such as those relating to
the reionization history of the Universe, do point to an early population that is more
massive than a standard Pop I or II distribution.  
Here, we will explore the consequences of an early population of intermediate mass (IM)
(2 - 8 M$_\odot$) stars, which dominate at a redshift $z \sim  10$.

While there are many theoretical arguments pointing to very massive
stars dominating Pop III \citep{bromm,bromm2}, there exist arguments pointing
to a first generation of stars with an initial mass function (IMF) peaked around
the intermediate mass range.  For example, the metal-free IMF predicted from opacity-limited 
fragmentation theory would peak around 4 -- 10 M$_\odot$ with steep declines
at both larger and smaller masses \citep{yoshii}.  
Primordial CMB regulated-star formation  also leads to the
production of a population of early intermediate mass stars  at low metallicity (Smith et al. 2009).

On the observational side, there is abundant evidence for an early contribution from IM stars. 
Intermediate mass stars have particular effects on chemical abundance
patterns.  Unlike their massive counterparts that end as core-collapse supernovae,
these stars produce very little in the way of heavy elements (oxygen and above),
but produce significant amounts of carbon and/or nitrogen and above all helium.
Indeed, there is evidence that the number of carbon enhanced stars
increases at low iron abundances \citep{rossi} necessitating a Pop III
source of carbon, possibly in the asymptotic giant branch (AGB) phase of 
IM stars \citep{fujimoto,aoki2,lucatello}.
To account for the large [C,N/Fe] ratios found in a large number of 
extremely iron-poor stars in our Galaxy, it was argued that an IMF peaked at
4 - 10 M$_\odot$ is needed \citep{abia}.
Furthermore, the presence of s-process elements, particularly Pb
at very low metallicity also points to AGB enrichment very early on \citep{aoki,sivarani}.
As a result, an early population of intermediate mass stars may
lead to a plausible explanation of recently observed carbon enriched iron-poor stars.

Isotopic abundances of C and Mg also lend clues to the early nucleosynthetic history
of the Galaxy.  Isotopic studies of the latter show a considerable amount of dispersion
at low metallicity ranging from low values of $^{25,26}$Mg/$^{24}$Mg, consistent
with pollution from core collapse supernovae, to high values of the Mg ratios
indicating the presence of AGB production of the heavy Mg isotopes. 
Core collapse supernovae produce almost exclusively $^{24}$Mg, 
while the observations of  \citet{yong} show enhancements
(relative to predictions based on standard chemical evolution models)
in both $^{25,26}$Mg. AGB stars were also concluded to be the source
of the high $^{25,26}$Mg/$^{24}$Mg ratios seen in the globular cluster NGC 6752
with [Fe/H] = -1.62 \citep{yong2}. A similar conclusion was drawn in  \citet{alibes,Fenner03}.

Determining the identity and nature of the first stars is a primary goal
in any attempt to understand the physics of the first billion years of the universe.
Cosmic chemical evolutionary models enable one to test many of the 
hypotheses for Pop III stars \citep{cf}. The bulk of the existing work in this area
considers either massive stars (M $ > 40$ M$_\odot$) or very massive stars (M $ > 100$ M$_\odot$).
However, given the indications discussed above for 
an early population of intermediate mass stars, it seems crucial to examine
the consequences for cosmic chemical evolution with IM Pop III stars.

The paper is organized as follows: In section \ref{mot}, we discuss some of the 
expected consequences 
for a population of intermediate mass stars.  These include a)
the effective prompt initial enrichment of helium, which we compare  to 
recent determinations of primordial helium \citep{it10,aos} that show an excess relative
to the predictions of big bang nucleosynthesis (BBN); b) the reionization of the universe, usually 
ascribed to very massive Pop III stars \citep{cen,hh,wl,bromm3}; c) the evolution of the IMF;
d) carbon-enhanced metal-poor stars; e) the density of white dwarfs; f) the rate of 
Type Ia supernovae at high redshift; and g) the evolution of N and Mg.

In section \ref{model1}, we introduce a relatively simple model of galactic
chemical evolution to highlight some of the (non-cosmological) effects of the IM
stars, particularly on the evolution of element abundances based on the work 
of \citet{foscv}.  This model
employs a bimodal star formation rate (SFR) which supplements a normal
IMF and SFR, with one enhanced with IM stars. In section \ref{model2},
we introduce a cosmic chemical evolutionary model 
based heavily on the hierarchical models developed in  \citet{daigne2} and \citet{rollinde}.
 We present 
our results in section \ref{results}. Discussion and concluding remarks are given in
section \ref{end}.

\section{Consequences of an early population of intermediate mass stars}
\label{mot}

\subsection{The primordial helium abundance}

Among the successes of the WMAP determination of cosmological parameters
is the accurate determination of the baryon density, $\Omega_B h^2$ or equivalently
the baryon-to-photon ratio $\eta \equiv n_{\rm b}/n_\gamma = 10^{-10} \eta_{10}$;
\citet{wmap,wmap10} find 
\beq
\label{eq:eta}
\eta_{10} = 6.19 \pm 0.15  \ \ .
\eeq
As a consequence,
it has become possible to treat standard big bang nucleosynthesis (BBN) 
as a zero-parameter theory \citep{cfo2,cfo3}, resulting in 
relatively precise predictions of the light element abundances of D, \he3, \he4, \li7 
\citep{cfo,coc,coc2,cyburt,coc3,cuoco,serpico,cfo5,coc10}. 

When compared with current observational determinations of the abundances of 
D, \he4, and \li7, we find varying degrees of concordance, none of which are perfect.  
Deuterium is typically credited as being responsible for the 
concordance between BBN and the cosmic microwave background determination 
of $\eta$.  At the current value of $\eta$ (eq.~\ref{eq:eta}),
the deuterium abundance is predicted to be D/H = $2.52 \pm 0.17 \times 10^{-5}$.
This can be compared with the abundance of deuterium
measured in quasar absorption systems.  The weighted mean value of
seven systems with reliable abundance determinations is  
D/H = $(2.82 \pm 0.21) \times 10^{-5}$ \citep[][and references therein]{pettini}.  However, 
the individual measurements
of D/H shows considerable scatter (a variance of 0.53)
 and it is likely that systematic errors dominate the uncertainties.
While the agreement is certainly reasonable, we do draw attention to the
fact that the predicted abundance is somewhat {\em lower} than the observed mean
and this (slight) discrepancy will serve as a constraint below.

The discrepancy between the predicted abundance of \li7 and the value found 
in halo dwarf stars is well documented \citep{cfo5}.
At the WMAP value of $\eta_{10}  = 6.19$,
\li7/H = $(5.12^{+0.71}_{-0.62}) \times 10^{-10}$
which is considerably higher than most observational determinations
\citep{spitex2,Spiteetal96,Ryan00,Asplundetal06,Bonifacioetal07,hos,hos2}.  For example, the recent analysis of \citet{sbordone}
finds \li7/H = $ (1.58 \pm 0.31) \times 10^{-10}$.
We comment below on the possible effects of astration in chemical evolution
models on the lithium abundance.

Among the light elements, helium has the most accurately predicted primordial abundance.  In addition, it is relatively insensitive to the baryon density and at 
$\eta = 6.19$, the helium mass fraction is 
\beq
Y_p = 0.2487 \pm 0.0002.
\label{ybbn}
\eeq
The standard BBN results obtained in   \citet{coc10} are similar to those
given above (Y = $0.2475 \pm 0.0004$,  D/H = $(2.68 \pm 0.15) \times 10^{-5}$,
 \li7/H = $ (5.14 \pm 0.50) \times 10^{-10}$).

On the observational side, the determination of the helium abundance in extragalactic HII
regions is fraught with difficulties \citep{os1}. Over the last 15 years,
improvements in the analysis has led to a systematically higher \he4 abundance.
\citet{itl94} and \citet{ppr00} introduced a `self-consistent' method for 
determining the abundance of \he4
simultaneously with the determination of the physical parameters associated with the HII
region.  They found $Y_p = 0.240 \pm 0.005$ and $0.235 \pm 0.002$ respectively.
Using the Monte Carlo methods described in \citet{os1} and selected
high quality data from \citet{itl94}, \citet{os2} found a higher value with
considerably larger uncertainties, $Y_p = 0.249 \pm 0.009$. 

Further improvements to the analysis were aided by new atomic data \citep{pfm} used 
in \citet{plp07} and \citet{its07} where higher helium mass fractions were obtained
$0.2477 \pm  0.0029$  and $0.2516 \pm 0.0011$, respectively, based on the new
He I emissivities and the collisional excitation of H I.  
The more recent work of \cite{it10} found a still higher value 
$Y_p = 0.2565 \pm 0.0010$(stat.) $\pm 0.0050$(syst.). 
Using the same relatively small set of high quality data from \citet{itl94},
\citet{aos} found a similarly high value of $Y_p = 0.2561 \pm 0.0108$.
The latter was a combined and uniform treatment of both hydrogen and helium
emission lines, the inclusion of the new He emissivities, wavelength-dependent 
effects of underlying absorption, and neutral hydrogen collisional corrections.
A combination of strong degeneracies leading to large uncertainties in the individual
helium abundance determinations found in an unrestricted Monte Carlo analysis
as well as the relatively short baseline used
in the regression with respect to O/H, the large uncertainty of 0.0108
makes certain that this determination is consistent with the BBN prediction at the 
WMAP value of $\eta$. Given, the short baseline of this data set, 
it may be argued that a weighted mean of the helium results is warranted rather
than the commonly used regression of Y vs O/H as in the case of all results quoted above.
The weighted mean found in \citet{aos} is 
\beq
Y_p = 0.2566 \pm 0.0028,
\label{yaos}
\eeq
and one could be tempted to claim a slight discrepancy between the
predicted and derived \he4 abundance.

If the central value of $Y_p$ is indeed 0.256, rather than invoking
non-standard BBN or a non-standard cosmological history,
it may be possible to argue for a prompt initial enrichment of He
by an early population of stars.  While the helium data responsible 
for $Y_p$ is at relatively low metallicity, the most metal-poor HII
region used in the analysis leading to Eq. (\ref{yaos}) is about 1/20
of solar metallicity. There is, therefore, a reasonable possibility that some helium
was produced early in the star formation history of pre-galactic structures.
Here, we consider specific models of chemical evolution
which may be responsible for such a prompt initial enrichment of \he4
remaining at the same time consistent with a multitude of evolutionary constraints.

\subsection{Reionization}


When WMAP first reported their result for a large optical depth
due to electron scattering \citep{wmap1}, which suggested a very early epoch
of reionization of the intergalactic medium (IGM), it had an immediate impact
on theories of cosmic chemical evolution.  The reported optical depth of 
$\tau_e = 0.17 \pm 0.04$ implied a redshift of reionization, $z_r \approx 20$. 
This put an immense amount of pressure on the models of the first stars,
and certainly pointed towards very massive stars as a potential solution
\citep{cen,hh,wl,bromm3,daigne1}.  In subsequent analyses by WMAP \citep{wmap10},
the optical depth
was lowered and currently it is reported to be $\tau_e = 0.088 \pm 0.015$, corresponding to 
$z_r = 10.5 \pm 1.2$.  This change is significant as 
it greatly relaxes the constraints imposed on the models.
First, reionization at the relatively late redshift of 11 allows more
time for structures and stars to form, and more importantly
at the later time, more baryons are incorporated in structures
(in the hierarchical structure formation picture), and this decreases the 
ionizing flux per star needed for reionization.

The early epoch of reionization is one of the 
driving forces behind the identification of Pop III stars with
supermassive stars.  However, it has been argued that the number of very massive 
Pop III stars needed for reionization would produce metal enrichments in conflict with
observations \citep{venk,daigne1}. On the other hand, it has been shown that 
a `normal' population of stars (i.e., a model with a standard SFR and Salpeter IMF)
does not produce enough ionization photons so as to reionize the IGM \citep{daigne1,daigne2}.
Because of the short timescales involved (to achieve reionization by a redshift of 11),
only stars with masses $M \ga 3$M$_\odot$ can contribute to reionizing the IGM by a redshift of 11.
While it has been shown that massive stars (in the range 40-100 M$_\odot$) are fully capable of
reionizing the IGM in a bimodal model of star formation \citep{daigne1,daigne2},
to the best of our knowledge the capabilities of IM stars have yet to explored in this context.
We will show below that indeed IM stars are capable of reionizing the IGM sufficiently rapidly.


\subsection{Evolution of the IMF}

One can make a phenomenological case from galaxy evolution  for an IM star-dominated IMF at $z\simgt 2.$ The arguments are  individually refutable but the cumulative effect is certainly suggestive. 

\begin{itemize}
\item 

Comparison of the cosmic star formation rate with the rate of stellar mass assembly,  inferred from integrating the luminosity function over various bands,
reveals a discrepancy at $z\simgt 2$ that increases towards higher $z$. The discrepancy is such that one {\it apparently} requires a reduced ratio of stellar mass to luminosity in star-forming galaxies. This is achievable if the IMF has an excess of intermediate mass  stars at these epochs.
No universal power law fit is acceptable at $z\simgt 0.5$ according to \citet{Wilkins1}.  A similar discussion also leads to a preferred SFR at $z\simgt 3$ that is $\sim 3$ times higher than the best fit to the stellar mass density \citep{Wilkins2}. There are other possible explanations, including uncertainties in stellar mass measurements and dust extinction, but no one of these is preferred.

\item 
Comparison of  the rate of luminosity evolution of massive early-type galaxies in clusters at $0.02<z<0.83$ to the rate of their color evolution requires that the IMF is weighted towards more massive stars at earlier epochs.  A specific  interpretation is that the characteristic stellar mass ($\sim 0.3 \rm M_\odot$ today)  had a value of $\sim 2\rm M_\odot$ at $z\sim 4$, \citep{Dokkum}.

\item
The evolution of the galaxy stellar mass--star formation rate relationship  for star-forming galaxies  constrains the stellar mass assembly histories of galaxies. Simulations including gas accretion reproduce a tight correlation but the evolution with redshift disagrees with the data.
The simulations generally prefer a constant specific star formation rate  
$\dot M_\ast/M_\ast$. 
The specific star formation rate is high at $z\simgt 2$, remains high to $z\sim 7$, but is   low today  \citep{Gonzalez}.
 To reconcile these results with the simulations, \citet{Dave 2008}  proposed that the stellar initial mass function (IMF) evolves towards more high-mass star formation at earlier epochs such that the critical stellar mass in the IMF varies approximately from the present  as $(1+z)^2$ to $z\sim 2$.

\end{itemize}

The arguments listed above apply to the high redshift universe where the dominant contributors to both light and stellar mass are massive star-bursting galaxies.  There is also some intriguing evidence from the local universe
that differentiates  the IMF in current epoch luminous  star-forming galaxies from their lower luminosity  counterparts.
 
Nearby massive star-forming galaxies seem to have an IMF that is more massive star-dominated than lower mass galaxies. There is evidence for a non-universal stellar initial mass function from the integrated properties of SDSS galaxies, according to \citet{Hoversten}. These authors  inferred the stellar initial mass function  from the integrated light properties of galaxies by comparing  galaxy H-alpha equivalent widths  (EW) with a broadband color index to constrain the IMF. 
They find that low luminosity galaxies have much lower EW(Hα) than expected for their colors and argue that this is due to the IMF in low luminosity galaxies having fewer massive stars, either by steeper slope or lower upper mass cutoff, than that for luminous galaxies.

A similar result for  HI-selected galaxies emerges via the  H-alpha and the far-ultraviolet continuum tracers of star formation. \citet{Meurer} find that the flux ratio H-alpha/FUV shows strong correlations with the surface-brightness in H-alpha and the R band: Low Surface Brightness (LSB) galaxies have lower ratios compared to High Surface Brightness galaxies.  The most plausible explanation for the correlations are systematic variations of the upper mass limit and/or slope of the IMF at the upper end. Massive galaxies in the local universe have an IMF that is preferentially enhanced in massive stars relative to low mass galaxies.

It seems that a case can be made that galaxies which form stars rapidly, specifically massive galaxies at high redshift and their nearby counterparts, possess an IMF that is relatively top-heavy (or bottom-light). We can speculate that essentially all vigorously star-forming systems at high redshift may similarly possess an IMF enriched in intermediate mass stars. These at least are the building blocks that in our model provide the sources of the bulk of ionizing photons at $z$ as high as $\sim 10$ and play a role in the recycling of processed gas.

\subsection{CEMPs and CRUMPs and the transition discriminant $D_{\rm trans}$}

 It is well known that  massive Pop III stars have a very specific impact on the
nucleosynthesis budget of the early Universe.
 In this context, it  is useful to incorporate  stellar halo observations, specifically from 
extremely metal-poor stars (EMPS, [Fe/H] $< -3$), ultra-metal-poor stars (UMPS, [Fe/H] $< -4$) and hyper-metal poor stars (HMPS, [Fe/H] $<  -5$) to derive constraints on the mass range of massive Pop III stars. 
From the nucleosynthetic point of view, halo stars have long been used to
constrain galactic chemical evolution models, but were, until recently, 
quite disconnected from cosmological
models. Indeed, these old low mass stars preserve their pristine composition 
on their surface. The abundance patterns seen in these metal-poor stars, originating presumably 
from  Pop III stars, can yield very precious information about the first nucleosynthesis processes in the Universe (subsequent to the Big Bang), and thus gives one the possibility of exploring the Universe during the period of reionization \citep{frebel07, frebel10}.

On the other hand, it has been shown \citep{Brommloeb, frebel07} that  the abundances of ionized carbon and neutral atomic oxygen are important for the transition from Pop III to
Pop II/I star formation. \citet{Brommloeb}  defined a  transition discriminant, 
\beq
D_{\rm trans} =  \log(10^{[C/H]}+0.3 \times10^{[O/H]})
\label{dtrans}
\eeq
which clearly reveals the nucleosynthetic imprint of Pop III stars. This function has the great advantage of being able to connect aspects of cosmological evolution with the C/O abundances in low mass stars. In this context, carbon-enhanced metal-poor stars or carbon-rich ultra-metal-poor stars 
(CEMPS and CRUMPS) seem to indicate an overabundance of carbon (and oxygen) 
at early stages of  evolution. 
CEMPS represent about 20 percent of the metal poor stars. Recent studies considering multi-zone chemical evolution models show that only models with increased C yields or an initial top-heavy IMF in the IM star mass range provide good fits of observations in the Milky Way \citep{mattsson10}. Note that these models also predict a nitrogen enhancement which will be considered below. 
Therefore, from the nucleosynthesis viewpoint it is interesting to consider IM stars as Pop III candidates, since they are producers of both C and He. 

Finally, observations of metal absorption features in the intergalactic medium (IGM) give important constraints on models of the formation and evolution of the earliest structures. Recent measurements of C in the IGM \citep{kramer10} show an increase of this element at about $z = 6$.
This seems to clearly indicate an increase in the star formation rate at this redshift able to produce carbon. Again,  IM stars are  good candidates for explaining this trend.
Note that we also predict that IGM material could have a high C/O ratio; indeed, we can obtain insight into this ratio from observations, though presently, it is difficult to obtain reliable C/O ratios from ionized abundance data (namely CIV and OVI).

\subsection{Element Abundances: N and Mg}

In the course of standard stellar evolution, nitrogen is produced through He burning. 
Primary nitrogen is also synthesized in massive stars in the H burning layer via fresh 
carbon coming from the He burning core. 
In intermediate-mass stars, nitrogen is produced together with carbon. 
Hot bottom-burning plays a dominant role for the synthesis of N and the amount of nitrogen
produced depends on the efficiency and the duration of nuclear burning. At lower metallicity, the production of nitrogen is favored over carbon. There are indeed systems at low metallicity,
such as 47 Tuc and M71, which show large nitrogen enhancements \citep{harbeck,briley}.
Abundance patterns found in these systems make it difficult to argue that the nitrogen
enhancements are due to internal contamination, and lend credence to the hypothesis
of an early IM stellar population \citep{amo}.

The evolution of the nitrogen abundance is very sensitive to specific parameter choices
in the chemical evolution model \citep{amo}. Included among these are the 
uncertainties stemming from the yields \citep{karakas07} particularly in the upper end of the IM 
mass range ($ > $ 3 M$_\odot$).
Unfortunately, this element is difficult to observe. Indeed, the best measurements 
come from the CN BX electronic transition band which is not
 generally detected. In this context, it is difficult to confront the calculated evolution of this element with available observations.
However, it is interesting to note that N is observed in the few CEMPS as shown below. 
 
 Magnesium is produced predominantly in Type II supernovae where it is formed in the 
 carbon and neon burning shells in massive stars. The dominant isotope formed in
 SNII is \mg24. The neutron-rich isotopes \mg25 and \mg26 are produced in the outer 
 carbon layer through $\alpha$ capture on neon. The yield of the heavier isotopes scales
 with the metallicity in the carbon layer and as a result, very little of these isotopes are produced
 at low metallicity. In contrast, significant amounts of \mg2{5,26} are produced during hot bottom-burning in the asymptotic giant branch phase in IM stars \citep{booth}.  These stars are hot enough
 for efficient proton capture processes on Mg leading to Al (which decays to the heavier Mg isotopes).
 The neutron-rich isotopes are also produced during thermal pulses of the helium burning shell. 
 Here, $\alpha$ capture on $^{22}$Ne (which is produced from $\alpha$ capture on $^{14}$N)
 lead to both \mg25 and \mg26. The latter process is very important for $\approx 3$ M$_\odot$
 \citep{karakas03}. 
 Several observations \citep{Shetrone,yong,yong2} indicate enhancements of the
 neutron-rich isotopes in low metallicity stars which could necessitate the presence of an
 early population of IM stars \citep{alibes,Fenner03,amo}.  
 
 Finally, it is interesting to note that Mg has been observed in ultra-faint dwarfs galaxies \citep{frebel10b} which are representative of the nucleosynthetic processes in the first galaxies. Specifically, in spite of the uncertainties related to the Mg data in these objects, we also note that  [Mg/Fe]  ratio is somewhat higher (about 1)  in these structures compared to the local data (about 0.3) in figure 1 of \cite{frebel10b}. Once again this could be a signature of the presence of IM stars in these primitive galaxies.

\subsection{White Dwarfs}

The best current limits on the halo old white dwarf population come from a combination of 
interpreting the MACHO and EROS microlensing events with improved white dwarf atmosphere modeling \citep{torres10}. The upper limit, including white dwarfs with hydrogen-deficient atmospheres, allows an increased microlensing optical depth. The upper limit to the contribution of halo white dwarfs is comparable to that of the thick disk, and accounts for at most 10\% of the halo dark matter between the MWG and the LMC. The MACHO upper limit, which has not been affected by subsequent microlensing studies,  is approximately 20\% (\cite{alcock00}). For comparison, our IM Pop III component gives a white dwarf mass fraction very comparable to the standard one ; it differs only by 15 {\%} (see Figure \ref{fig:Mass} below).

\subsection{Supernova rates}
\label{presn}

Among the important constraints imposed on the models of cosmic chemical evolution
are the rates of supernovae (SN). These are intimately linked to the
choice of an IMF and SFR and represent an independent probe of the
star-forming universe. While Type II rates directly trace the star formation history,
there is a model-dependent time delay between the formation (and lifetime) of the 
progenitor star and the Type Ia explosion.  Since existing data is available only for relatively low
redshifts, we discuss the predicted SN rates in the context of three different models and give some constraints related to the SNIa rate coming from the IM star mode. 

Our early IM stellar population hypothesis inevitably produces many SN at high redshift. The predicted SFR and SN rates are presented below. The SFR rate corrections (for dust, luminosity function etc)  are notoriously unreliable at $z\simgt 8$ where our IM population  (predominantly formed earlier than  $z\sim 10$) makes a significant  contribution. The SN Ia rates provide a potentially more robust  discriminant.  However we note here that three issues help reconcile our SN rate predictions with observations of SNIa at $z\simgt 1.$    Most importantly, the selection efficiently of SNIa, in the absence of near infrared selection, dives rapidly to zero at high redshift (beyond $z\sim 1.5$). Indeed, from \citet{perrett10}, the extrapolation of the efficiency to $z=2$ implies a steep drop (to only a fraction of a percent). Moreover the SNIa precursors are predominantly in low mass galaxies, which contain sub-luminous SNIa, by as much as $0.5$ magnitude
 \citep{sullivan10}.   Thirdly, it may be that at low metallicity ([Fe/H] $ < -1.1$), the absence of iron lines
 driving optically thick winds may greatly inhibit the number of SNIa \citep{nomoto}.
 Finally, note that there are large uncertainties related to the time delay between IM star
formation and the resulting SNIa explosion.  
These uncertainties add to the complexity of the question, 
since the predicted bulk of SNIa can be found at higher or lower high redshift,
according to the value of this time delay  \citep{totani10}.

\section{Standard chemical evolution model}
\label{model1}

There are only a few basic ingredients to a standard (non-cosmological) galactic evolution model.
These include: the specification of an IMF, usually assumed to be a power law
of the form $\phi(m) \sim m^{-x}$; a star formation rate, which may be proportional
to some power of the gas mass fraction or a specified function of time;  
a specification of stellar properties such as lifetimes and elemental yields.  
More complex models will include  infall of gas from the IGM, and outflows 
from stellar induced winds. Certain observations such as the metallicity distribution in 
low mass stars may call for the prompt initial enrichment of 
metals which can be accomplished using bi-modal
models of chemical evolution in which a standard mode (e.g. a Salpeter IMF 
and a SFR proportional to the gas fraction) is supplemented with
a shallow IMF providing predominantly massive stars over a brief period of time (or metallicity).
In this section, we will show that
this approach cannot work for the prompt enrichment of \he4
as the most massive stars would produce heavy elements very early and thus 
would not enhance the ratio of He to metals. In other words, the enrichment of 
helium would occur too late and would not explain the enhancement in low metallicity regions.

In contrast, bi-modal models which are augmented with a population of intermediate mass
stars (with mass between 2 and 8 M$_\odot$), can indeed provide an enrichment
of helium at low metallicity in simple galactic chemical evolution models.  
While the model presented in this section cannot directly be extended to 
a model of cosmic chemical evolution, it does provide some insight to the 
needed bimodality involving IM stars that will carry over into a more sophisticated model
based on hierarchical structure formation as in  \cite{daigne2}.

In \cite{foscv},
we constructed chemical evolution models
which led to significant D astration at low metallicity
and also had significant white dwarf production.
These models were based on a bimodal construction
where the `massive' mode consisted of IM stars.
Here we consider the effect of such models on 
\he4 production at low metallicity.

We illustrate results with the simple closed-box model
of \citet{foscv},
which neglects both infall and outflow of baryons.
Baryonic cycling through stars is determined
by the creation function $C(m,t)$
\beq
{\rm d}N_\star =  C(m,t) \; {\rm d}m \; {\rm d}t ,
\eeq
which describes the mass distribution of new stars
and the total star-formation rate $\Psi(t) = \int C(m,t) \ dm$.
We adopt a
``bimodal'' 
creation function 
\beq
\label{eq:creation}
C(m,t) = \psi_1(t) \phi_1(m) + \psi_2(t) \phi_2(m) ,
\eeq
each term of which represents a distinct star-formation mode.

In the first term of eq.~(\ref{eq:creation}) we encode
an early burst of IM star formation; to do this we adopt 
the star formation rate 
\beq
\label{eq:burst}
\psi_1 = x_{\rm burst} \frac{M_{\rm B}}{\tau_1} \ e^{-t/\tau_1} ,
\eeq
with a burst timescale  $\tau_1 = 100$ Myr.
The parameter $x_{\rm burst}$ controls the fraction of
the baryonic mass $M_{\rm b}$ which is 
processed in the burst; 
we will show results for different values of this parameter,
but clearly we must have $x_{\rm burst} \le 1$.
Since our focus is on IM stars, 
the IMF for this mode is a power-law (Salpeter) form:
$\phi(m) \propto m^{-2.3}$, for $m \in [2\msol,8\msol]$.

The second term in eq.~(\ref{eq:creation}) describes
normal star formation.
Again, following \citet{foscv},
we adopt the classic form $\psi_2 = \lambda g(t) M_{\rm gas}$, 
with $\lambda = 0.3 \; {\rm Gyr}^{-1}$, and
a smoothing factor $g(t) = 1-e^{-t/0.5 {\rm Gyr}}$
which ensures that star formation begins in the bursting IM
mode, which then smoothly goes over to the standard mode.
The normal-mode IMF is of the Salpeter form $\phi_2(m) \propto m^{-2.6}$,
but now with $m \in [0.1\msol,100\msol]$.
Both IMFs are normalized
in the usual way $\int m \ \phi_2(m) \ dm = 1$.

\begin{figure}[htb]
\begin{center}
\epsfig{file=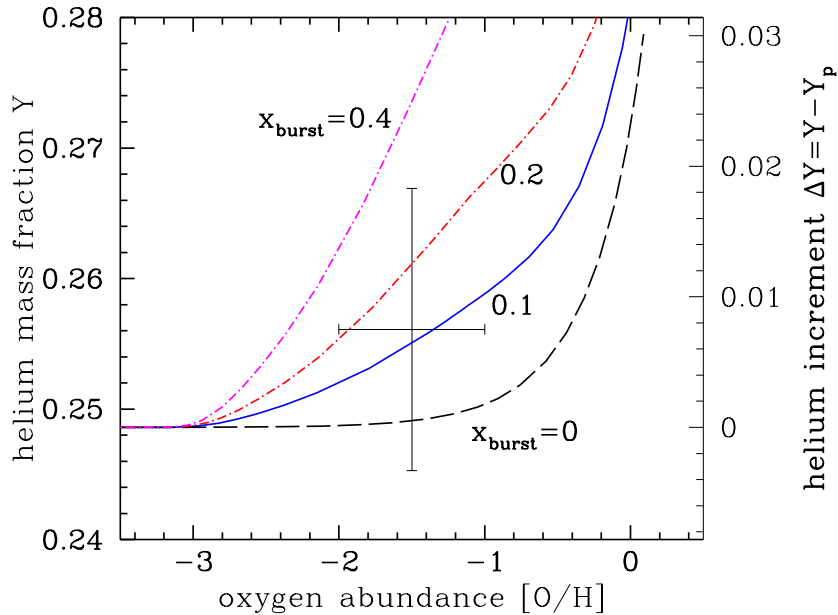, height=5.0in,angle=270}
\end{center}
\caption{The \he4 mass fraction in several bimodal models of chemical 
evolution.  Curves show different fractions $x_{\rm burst}$ (eq.~\ref{eq:burst}) of baryons processed
through the burst phase.
The data point is representative of the helium data in low metallicity 
extragalactic HII regions. 
\label{fig:simpHe}
}
\end{figure}

In Fig. \ref{fig:simpHe}, we show the evolution of the \he4 mass fraction, $Y$
as a function of the oxygen abundance (relative to solar oxygen). 
Curves are shown for different values of the burst mass fraction 
$x_{\rm burst}$ in eq.~(\ref{eq:burst}).
In the case of normal star formation (i.e., when $x_{\rm burst} = 0$ and
the IM mode is absent), as O increases we see He rise from its
primordial value, but undershoots the low-metallicity data.
That normal star formation fails to match the data traces back to
nucleosynthesis patterns in high-mass stars, which dominate
the production of O and of new He in this scenario.
Specifically, for high-mass stars, the ejecta have
$\Delta Y/\Delta Z_{\rm O} \sim 1$, i.e., comparable masses
of new He and of O \citep{ww95}.
Thus at low metallicity $Z_{\rm O} \ll Z_{\rm O,\odot}  \sim 0.01$, 
we have $\Delta Y \ll 0.01$, whereas the data seem to require
$\Delta Y_{\rm obs} \simeq 0.01$.  
This not only explains the problem with this simple scenario, but
also suggests the solution:  invoke a stellar population which
has substantially different helium and metal nucleosynthesis.

Indeed, Fig.~\ref{fig:simpHe} shows
that as we turn on the bursting IM mode via $x_{\rm burst} > 0$, 
there is an abrupt rise in He at low O abundances,
with a larger rise for larger $x_{\rm burst}$.
As anticipated, this rise now allows for the predictions to match
the data, with good fits for $x_{\rm burst} \sim 10\%-20\%$.
The origin of this rise is clearly reflects the nucleosynthesis
patterns of IM stars, whose ejecta are much more enriched in
new He than in O, relative to massive stars.
Specifically, for IM stars at low metallicity
in the  $m = (4\msol,8\msol)$ we have 
$\Delta Y/\Delta Z_{\rm O} \sim (50,300)$ \citep{hoek97}.
These large ratios trace back to the very small O production
in these stars.

The results thus far are for the simplest possible model:
a single-zone closed box.
Staying with single-zone models, even if we include
infall and/or outflow the results should be similar.
In the case of primordial infall, both the metal and new He
yields will be diluted, and by the same factor.
Thus the He-O trends should stay the same.
Similarly, in the case of outflows, the abundances in the gas remaining
in the box are only affected if the flow ejects
one element preferentially relative to the other.  But
again, given that the O and new He  are produced in the same environments,
any physically realizable winds will remove them
together, leaving the remaining gas with He-O trends unaffected.

The lessons we draw from our simple model 
are that an increase in helium at low metallicity,
as suggested by observations,
can be accomplished at the cost of
processing 10\%--20\% of observable cosmic baryons through IM
stars.
But as shown in \citet{foscv} and discussed above, 
such a scenario also has substantial consequences, including 
other element abundances, white dwarfs, and Type Ia supernovae.
To examine all of these issues in a more realistic manner 
requires a chemical evolution model in a cosmological context.

\section{A cosmological chemical evolution model}
\label{model2}

\subsection{Generalities}
The study of star formation in a cosmological context requires the inclusion of i) a model of 
dark matter structure formation, and ii)  accretion and outflow of baryonic matter with respect to existing and forming structures.

We have developed a detailed model of cosmological chemical
evolution \citep{daigne2,rollinde} using a description of non linear structures, based on the standard
Press-Schechter (PS) formalism  \citep{ps}. 
Baryons are placed  1) within stars or their remnants within collapsed structures,
2) in gas within collapsed structures (the interstellar medium (ISM)), or 
3) outside of 
structures (the intergalactic medium,  IGM). 
The rate at which structures accrete mass is determined by
a Press-Schechter distribution function, $f_{PS}(M,z)$. The model included mass (baryon)
exchange between the IGM and ISM, and between the ISM and stellar component.  
In this study we assume that the minimum mass of dark matter
halos for star-forming structures is $10^7$ M$_\odot$. 

Once the model is specified, we can follow many astrophysical quantities such as:
 the global SFR, the optical depth, the supernova (SN)  rates and the abundances of individual elements (Y, D, Li, Fe, C, N, O, Mg abundances) which are used to tackle specific questions related to early star formation and the reionization epoch.
In this study,  we consider two distinct modes of star formation: a normal mode of Pop II/I and  a massive or intermediate mass mode of Pop III stars. 

Throughout this paper, a primordial power spectrum with a power law index $n=1$ is assumed
and we adopt the cosmological parameters of the so-called concordance model
\citep{wmap10}, i.e. $\Omega_m=0.27$, $\Omega_\Lambda=0.73$, $h=0.71$ and
$\sigma_8=0.9$.

\subsection{Star formation rate scenarios}

As in the simple model described in section \ref{model1}, the models considered are bimodal. 
 Each model contains a normal mode with stellar masses
between 0.1 M$_{\odot}$ and 100 M$_{\odot}$, with a near Salpeter IMF ($x = 1.6$) 
(hereafter called model 2).

We fit the SFR history of Pop II/I stars to the data compiled in \citet{hb06} (from $z=0$ to
5), and to the recent measurements at high redshift by \citet{bouwens10} and \citet{gonzalez10}.
These observations place strong
constraints on the Pop II/I SFR.
The fit for the SFR, $\psi(z)$, is based on the mathematical form proposed
 in \cite{sp03} 
 
\begin{equation}
\psi(z) = \nu\frac{a\exp(b\,(z-z_m))}{a-b+b\exp(a\,(z-z_m))}\,.
\end{equation}
The amplitude (astration rate) and the redshift
 of the SFR maximum are given by $\nu$ and  $z_m$ respectively, while
$b$ and $b-a$ are related to its slope at low and high redshifts
respectively. The normal mode is fitted using: $\nu_{\rm II/I}=0.3$ M$_\odot$ yr$^{-1}$ Mpc$^{-3}$, 
$z_{\rm m\, II/I}=2.6$, $a_{\rm II/I}=1.9$ and $b_{\rm II/I}=1.1$.
The SFR of this mode peaks at $z \approx 3$.  

In addition to the normal mode, we add a mode for Pop III stars.
Our primary interest here is a model for IM Pop III stars,
and we choose an IM component of stars with 2 - 8 M$_\odot$ (hereafter model 3),
assuming an IMF with slope 1.3.
Some fraction of the IM stars become SNIa.  
The SFR parameters are the following: $\nu_{\rm IIIa}=3$ M$_\odot$ yr$^{-1}$ Mpc$^{-3}$, 
$z_{\rm m\, IIIa}=10$, $a_{\rm IIIa}=1.9$ and $b_{\rm IIIa}=1.1$. 
The redshift peak is chosen so that the IM stars are born in time to affect low-metallicity abundances, and the normalization 
is chosen to allow a substantial effect on \he4.

For comparative purposes, we also consider a massive component with stars between 36 - 100
M$_\odot$ with an IMF slope of 1.6, which terminate as Type II supernovae (SNII) (hereafter model 1). 
The SFR parameters are the following: 
$\nu_{\rm IIIb}=0.0016$ M$_\odot$ yr$^{-1}$ Mpc$^{-3}$, 
$z_{\rm m\, IIIb}=22.8$, $a_{\rm IIIb}=4$ and $b_{\rm IIIb}=3.3$.
For a detailed description of the model see \citet{rollinde} and \citet{daigne2}.
Note that both models 1 (massive) and 3 (IM) include the normal mode (model 2).

\begin{figure}[htb!]
\begin{center}
\epsfig{file=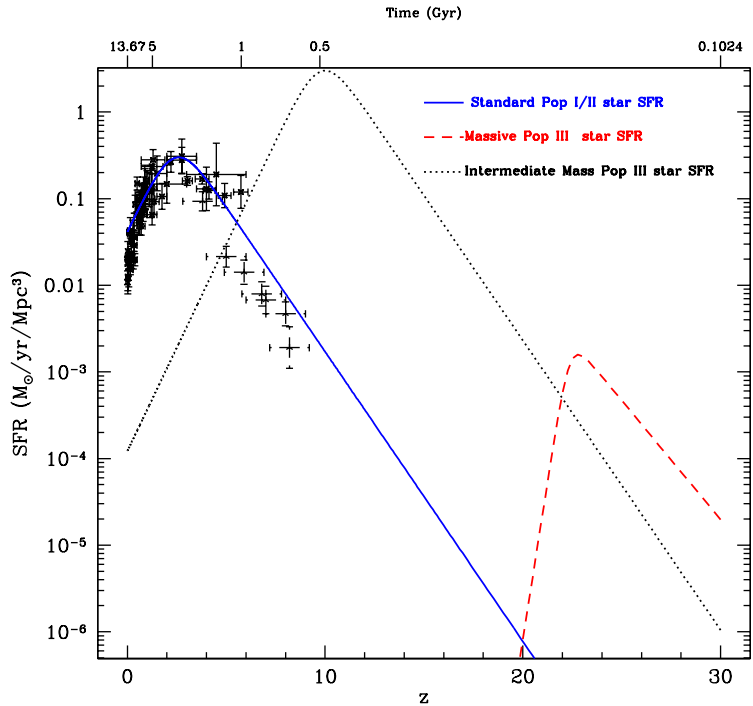, height=4.0in}
\end{center}
\caption{The cosmic star formation rate (SFR) as a function of redshift. 
The data (solid black  points) are taken from \cite{hb06}. Dashed black points come from \cite{bouwens10} and from \cite{gonzalez10}.
The blue solid line represents the standard SFR with a Salpeter IMF and a mass range: 0.1 $<$ M/M$_\odot <$100.
The dashed red line represents the massive Pop III stellar mode, with a Salpeter IMF and a mass range: 36 $<$ M/M$_\odot<$ 100. The dotted black curve represents the IM SFR mode, 
with a Salpeter IMF and a mass range: 2 $<$ M/M$_\odot<$8.
\label{fig:sfr}
}
\end{figure}

In Fig. \ref{fig:sfr}, we show the adopted SFR for each of the three cases considered.
As one can see, the normal mode is chosen to fit the observations also plotted in the 
figure.  As the data extend only up to $z \approx 8$, there is essentially no constraint on
the massive mode which dominates at $z \ga 20$.  
Data at $z\simgt 8$  is highly uncertain due to unknown systematics involving, among other effects, the dust corrections and adopted rest-frame UV luminosity function, and presents essentially no conflict with our models (\cite{labbe10}).

Fig. \ref{fig:sfr} also shows that the IM mode represents a significant
processing of baryons.  The peak of the IM star formation rate is
about a factor $\sim 10$ higher than the peak of the
normal star formation rate, and a factor $\sim 10^3$ higher
than the massive Pop III star formation rate.
On the other hand, the IM star formation peak lasts for about
a factor $\sim 10$ less time than the normal star formation rate.
Thus we expect that roughly comparable amounts of
baryons will be processed through these two modes.
In contrast, a much smaller fraction
of baryons are processes through the massive Pop III mode.
One should recall, however, that in our model many baryons
always remain the IGM and never reside in galaxies nor are
processed though stars.

\begin{figure}[htb!]
\begin{center}
\epsfig{file=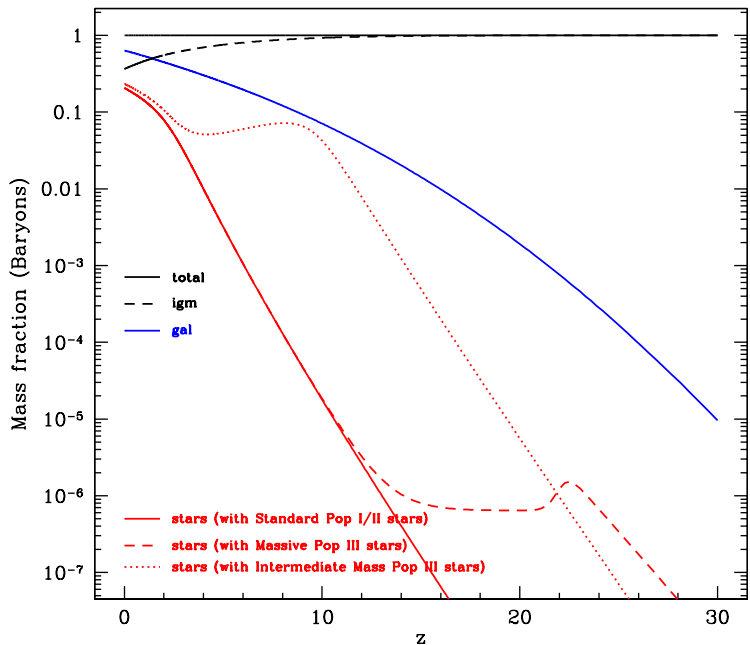, height=4.0in}
\end{center}
\caption{Different mass fractions as a function of redshift. The flat solid black line (top of the figure) represents the total baryons in the Universe. 
The IGM fraction is shown by the dashed  black curve, the galactic fraction is shown by the growing solid blue curve. The mass fraction of stars is plotted for the three models considered (red curves): solid line for model 2 (normal mode), dashed line for the model 1 (with massive Pop III stars) and the dotted line for  model 3 (with IM stars).
 \label{fig:Mass}
}
\end{figure}

Figure \ref{fig:Mass} represents the mass fraction of each component related to the total mass of baryons:
IGM, Galaxies and stars.
Note that the addition of the IM star component does not affect the mass fraction of stars (including remnants) at low redshift .

Note that the star fraction in the IM mode has an initial peak at
about $\sim 10\%$ of all baryons. This occurs at $z \sim 10$.  The
star fraction then drops as the IM stars die off, then rises with
normal star formation.  Thus we see that in our model, about 10\% of
the baryons are processed through the IM mode.  In contrast,
we see that the massive Pop III mode only processes about $\sim 10^{-6}$
of all baryons.

\subsection{Yields and lifetimes}

The lifetimes of intermediate mass stars ($0.9<M/\msun <8$) are taken
 from \citet{maeder89} and  from
\citet{schaerer02} for more massive stars. 
Old halo stars with masses below $\sim 0.9\ \msun$ have a lifetime  long enough to be
observed  today. They are assumed to inherit the abundances of the ISM at the time of their 
formation.
Thus, their observed abundances reflect, 
in a complex way (due to exchanges with the IGM),
the yields of all massive stars that have exploded earlier.

The yields of stars depend on their mass and their metallicity, but not
on their status (i.e. Pop II/I or Pop III). Some Pop II/I stars are
massive, although in only a very small proportion since we use a slightly steeper than Salpeter
IMF. Pop III stars are all massive stars. We use the tables of
yields (and remnant types)  given in~\citet{hoek97} for intermediate mass stars ($<8 \msun$),
and the tables in~\citet{ww95} for massive
stars ($8<M/\msun <40$ ). An interpolation is made between
different metallicities (Z=0 and Z=$10^{-4,\, -3,\, -2,\, -1,\, 0}$Z$_\odot$)
and we extrapolate the tabulated values beyond 40 \msun.

\section{Results}
\label{results}

Having specified the models under consideration, we can now systematically
consider the consequences of our particular choices for Pop III stars.
For most of the results which follow, we will compare three distinct model choices.
1) model 2 alone, ie, only Pop II/I stars; 2) model 3, a bimodal model
including the normal mode (model 2) plus the IM mode of Pop III stars ; 3) model 1, 
a bimodal model including the normal model (model 2) plus the massive mode of Pop III stars.

\subsection{Reionization}
Having set the SFR in our model, we now compute the electron scattering optical depth for our
choice of IMF. The evolution of the volume filling fraction of ionized regions is given by:

\beq
\frac{\mbox{d}Q_{{\rm ion}}(z)}{\mbox{d}z} = \frac{1}{n_{\rm
 b}}\frac{\mbox{d}n_{{\rm ion}}(z)}{\mbox{d}z}-\alpha_{{\rm B}}n_{{\rm b}}C(z)
   Q_{{\rm ion}}^{2}(z)\left(1+z\right)^{3}\left|\frac{\mbox{d}t}{\mbox{d}z}\right|\mbox{\ ,}
\eeq
where $n_{\rm b}$ is the comoving density in baryons, $n_{{\rm ion}}(z)$ the comoving density
of ionizing photons,  $\alpha_{{\rm B}}$ the recombination coefficient,  and $C(z)$ the
clumping factor. This factor is  taken from \citet{greif06} and  varies from a value of 2 at $z\leq20$ to a constant value of 10 for 
$z<6$. The
escape fraction,  $f_{\rm esc}$, is set to 0.2 for both Pop III and PopII/I . The number of ionizing photons  for massive
stars is calculated using the tables given in \cite{schaerer02}. Finally, the Thomson optical
depth is computed as in \cite{greif06}:
\begin{equation}
\tau =c\sigma_{{\rm T}}n_{\rm b} \int_{0}^{z}dz'\,Q_{{\rm ion}}(z')\left(1+z'\right)^{3}\left|\frac{\mbox{d}t}{\mbox{d}z'}\right|\mbox{\ ,}
\end{equation}
where $z$ is the redshift of emission, and $\sigma_{{\rm T}}$ the Thomson scattering
 cross-section.

\begin{figure}[htb!]
\begin{center}
\epsfig{file=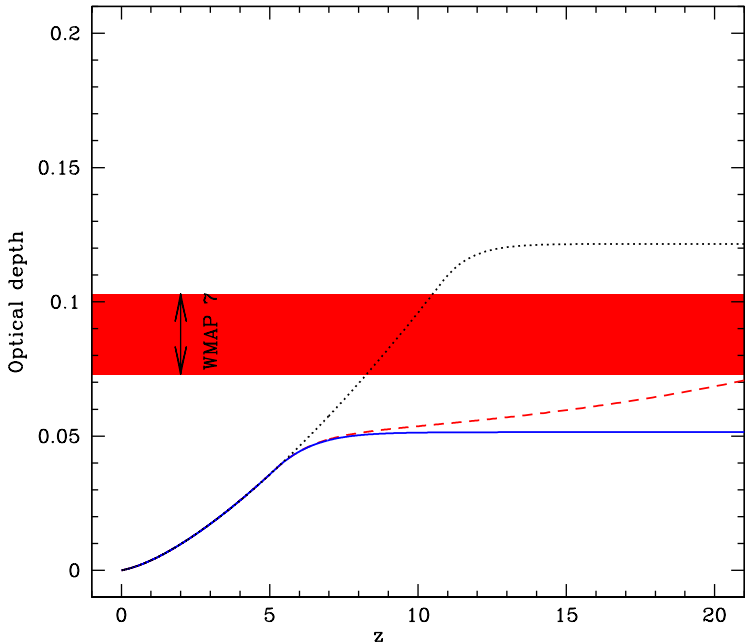, height=4.0in}
\end{center}
\caption{The optical depth as a function of the redshift.  The red range corresponds to the observed results from WMAP7( \cite{wmap10}).  
The red dashed line corresponds to model 1 (with Massive Pop III stars), the blue solid line to 
model 2 (the normal mode) and the black dotted line to  model 3 (with IM stars). 
 \label{fig:OpticalD}}
\end{figure}

In Figure \ref{fig:OpticalD}, we plot the
integrated optical depth from $z=0$ to $z$. The red band represents the observed results from WMAP7 \citep{wmap10}.  As one might expect, the normal mode alone (model 2 shown as the solid blue curve) is not capable of producing a sufficiently large optical depth.  The massive mode
(model 1 shown by the red dashed curve) does manage to reionize the IGM, but only barely.
Of course more reionization is possible by increasing the SFR for the massive mode, but this would
lead to complications in elemental abundances.
 We see that the IM mode  (model 3 - shown by the black dotted curve) is able to easily reionize the Universe compared to models 1 and 2.
Our models allow an escape fraction as low as $\sim 0.1-0.2,$ which provides considerable
flexibility in accounting for reionization. While the ratio of ionizing photons
from a 6\msol\ star vs that of a 20\msol\ star is small, one has many additional stars in the
IM scenario relative to the number in the massive mode.

\subsection {Nucleosynthesis evolution}

\label{sect:nuke}

We next consider the evolution of the abundances of He, CNO, Fe and Mg  as a function of the redshift
 and/or the metallicity. We also show the evolution of D and \li7 due to early astration.
 
\begin{figure}[htb!]
\begin{center}
\begin{tabular}{cc}
\epsfig{file=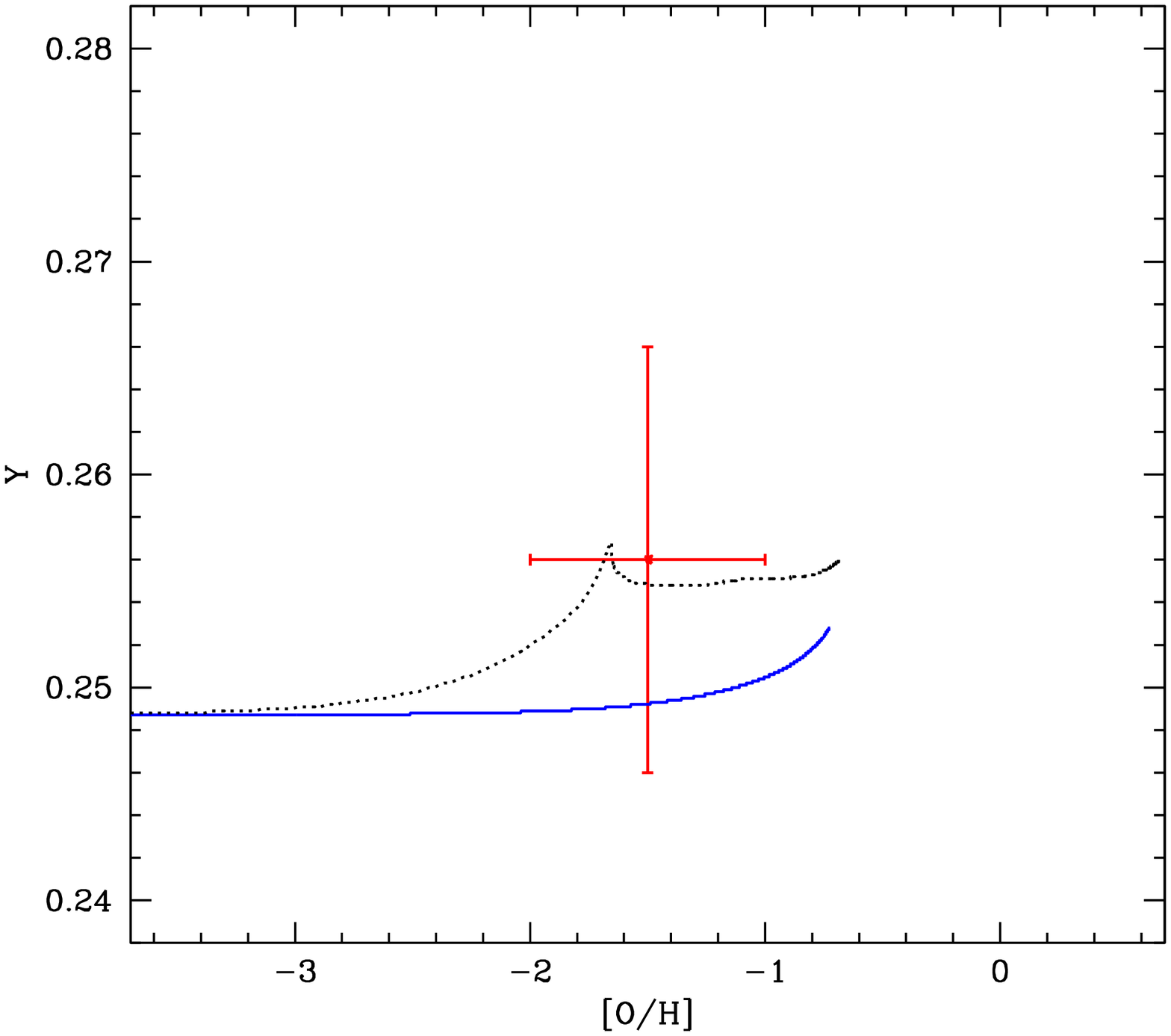, height=3.0in} &
\epsfig{file=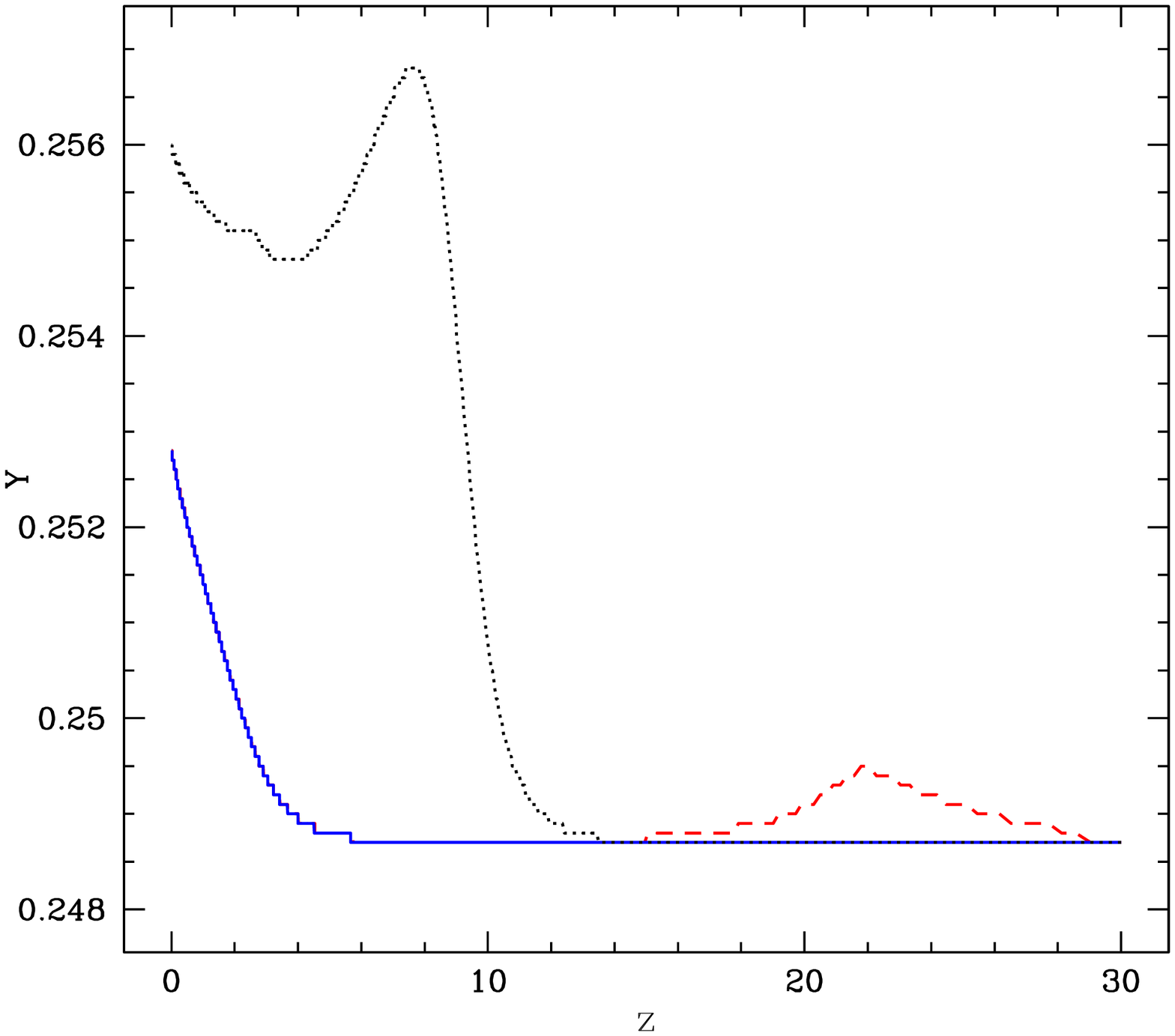, height=3.0in}\\
\end{tabular}
\end{center}
\caption{\textit{Left:} Evolution of the helium mass fraction as a function of [O/H] (relative to the solar value).
The solid blue line corresponds to the both models: 2 (standard model) and 1 (with a massive Pop III mode). The dotted black curve corresponds to  model 3 ( with IM Pop III stars). The red point comes from \cite{aos}.
\textit{Right:}
Evolution of the helium mass fraction as a function of redshift . The red dashed  line corresponds to the Massive Pop III model , the blue solid  line to the standard model, and the black dotted one to the IM Pop III model.
\label{fig:HEOz}
}
\end{figure}

We begin with the evolution of the \he4 abundance.
In the left panel of Fig. \ref{fig:HEOz}, we show the analogue of Fig. \ref{fig:simpHe} for the 
bimodal models based on hierarchical structure formation. Plotted is the  
\he4 mass fraction, $Y$, vs. [O/H].
The solid blue curve corresponds to 
the model with only a Pop II/I contribution. This result is indistinguishable from 
that produced by models including the massive Pop III contribution (not shown). 
This can be understood as the bulk of the \he4 and oxygen are derived from the same
stars in the two models. In both cases, the helium abundance
begins at the BBN primordial value and remains rather flat until the 
oxygen abundance is roughy 1/10th  of solar.  
In contrast, as shown by the black dotted curve, in the IM Pop III model, 
\he4 is produced early in IM stars
with little or no accompanied oxygen.  As a result, the \he4 mass fraction 
begins to grow at very low [O/H] and for the model parameters chosen,
plateaus at value close to $Y \approx 0.256$, in good agreement with recent
determinations \citep{it10,aos}.  As in the case of the simple model
discussed in section \ref{model1}, this model, therefore produces an effective prompt initial
enrichment of \he4 in low metallicity galaxies. Note however, that due to the size of the
error bar associated with the observations, we cannot exclude the possibility that
no enrichment occurred, leaving models 1 and 2 viable.

It is helpful to compare the helium--oxygen trend in Fig 5(a) 
with the simple closed box results in Fig. 1.  
The latter shows results for different fractions $x_{\rm burst}$ of
baryons processed through the IM mode.  As noted above,
our full hierarchical model cycles about 10\% of all baryons
through IM stars; thus Fig.~5(a) should be compared to the
$x_{\rm burst} = 0.1$ curve in Fig.~1.  Indeed, we see that both 
show very similar \he4 evolution.

In the right panel of Fig. \ref{fig:HEOz}, we show the corresponding evolution of the \he4 abundance
with redshift. The dashed red curve corresponds to the massive Pop III mode. 
Once again, the standard model and the massive Pop III model show nearly identical
histories.  We do see however, a modest increase in the helium mass fraction at high redshift
due to the massive Pop III mode. This is diluted by further infall as structures continue to grow.
In the IM Pop III model, there is a significant enhancement in the helium mass fraction around 
$z \sim 8$.  This too is slightly diluted with infall at later times.

\begin{figure}[htb!]
\begin{center}
\begin{tabular}{cc}
\epsfig{file=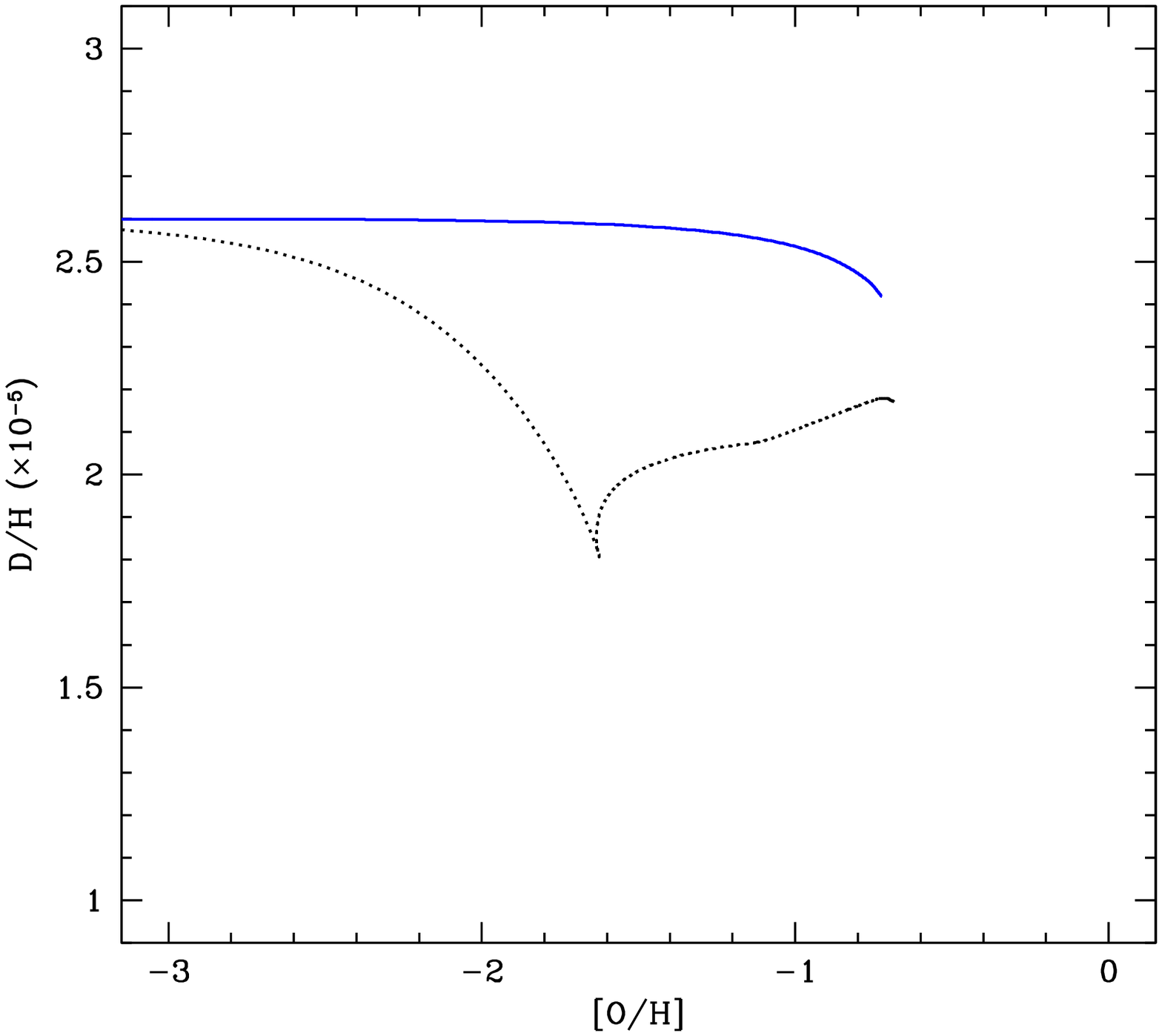, height=3.0in}&
\epsfig{file=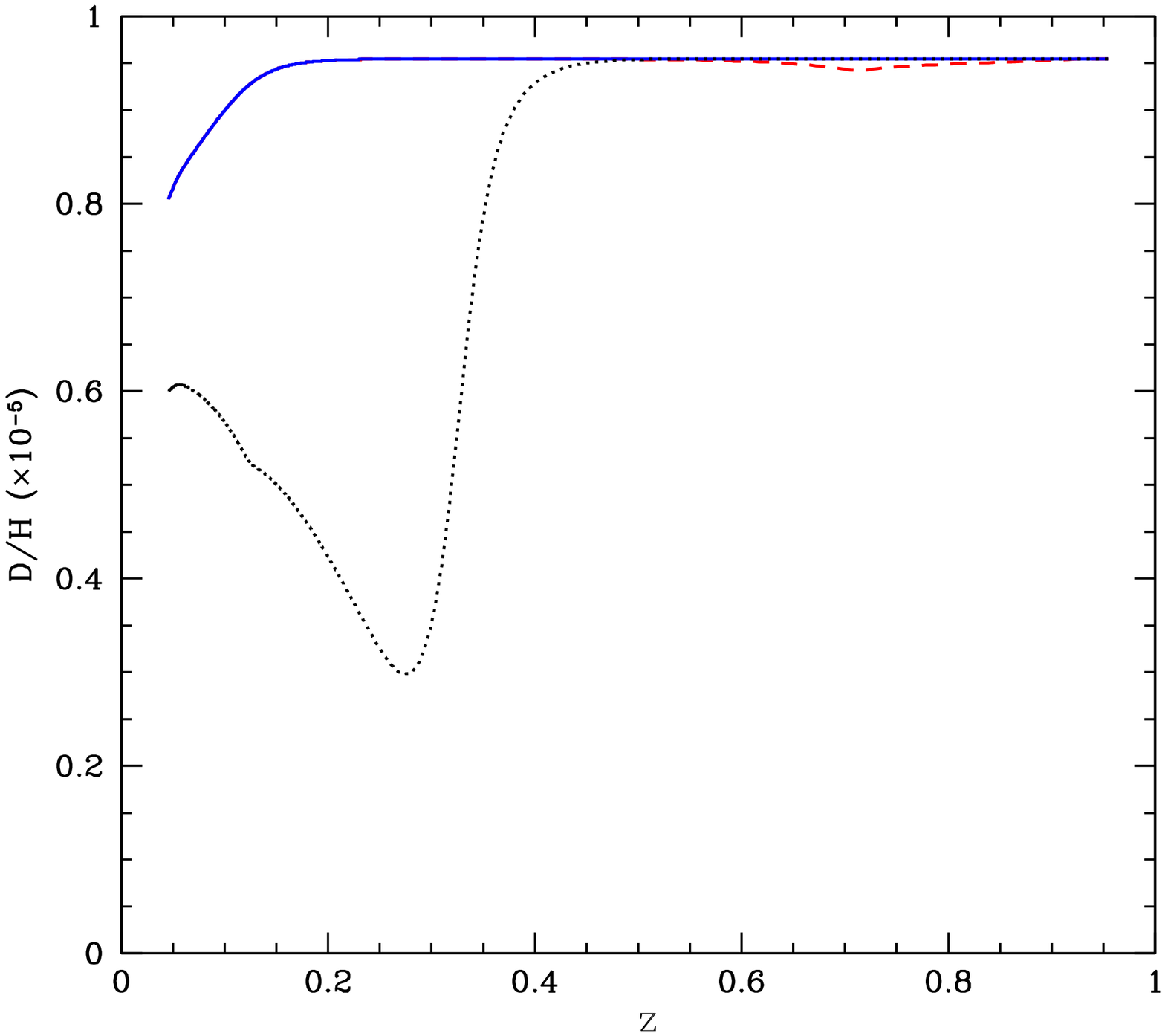, height=3.0in}\\
\end{tabular}
\end{center}
\caption{\textit{Left: }As in Fig. \ref{fig:HEOz} showing the evolution of the deuterium abundance  
as a function of [O/H], \textit{Right:} showing the evolution of the deuterium abundance with redshift.
Note that the vertical axis is offset from zero, 
in order to more clearly show the (relatively small)
effect on D. 
\label{fig:deutoz}
}
\end{figure}

\begin{figure}[htb!]
\begin{center}
\begin{tabular}{cc}
\epsfig{file=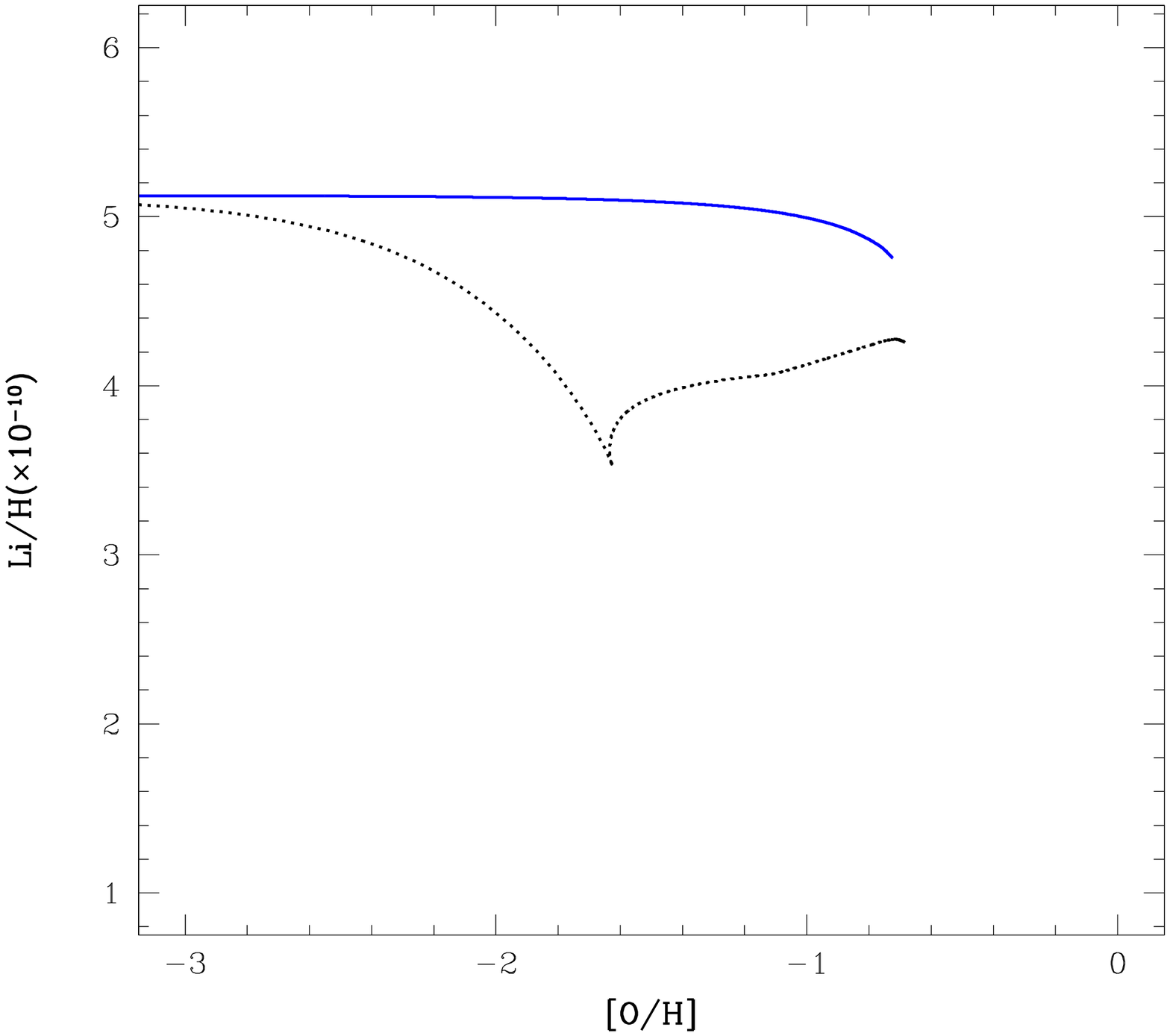, height=3.0in}&
\epsfig{file=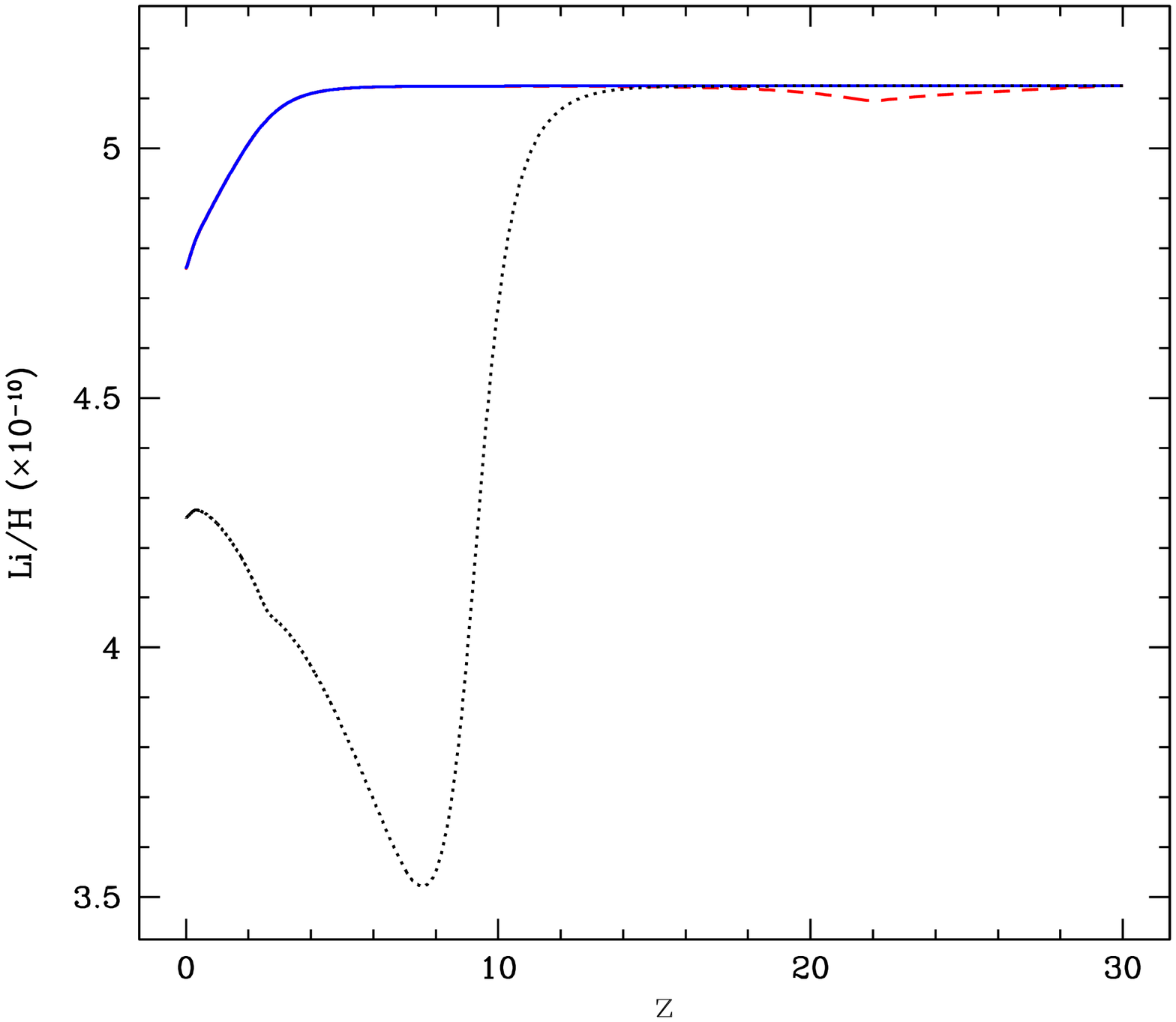, height=3.0in}\\
\end{tabular}
\end{center}
\caption{\textit{Left:}As in Fig. \ref{fig:HEOz} showing the evolution of the lithium abundance  
as a function of [O/H] . \textit{Right:} showing the evolution of the lithium abundance with redshift.
\label{fig:lioz}
}
\end{figure}

We also show in Figs. \ref{fig:deutoz} - \ref{fig:lioz} the corresponding
evolution for D and \li7; note that the vertical axes have zero offset in order
to more clearly distinguish the model predictions.
In each case, there is little difference between
the standard model and the massive Pop III model.  Intermediate mass stars on the other hand 
are known to deplete D/H \citep{foscv}. In this case, the degree of depletion is not severe (about 15\%)
but it does move the BBN value further away from the abundance determined in quasar 
absorption systems. The \li7 astration factor is the same as for D/H and in this case
moves the BBN value closer to the abundance determined in halo stars. However,
an abundance of $\sim 4 \times 10^{-10}$ is still very far from the observed plateau
value between 1 and 2 $\times 10^{-10}$. 

\begin{figure}[htb!]
\begin{center}
\begin{tabular}{cc}
\epsfig{file=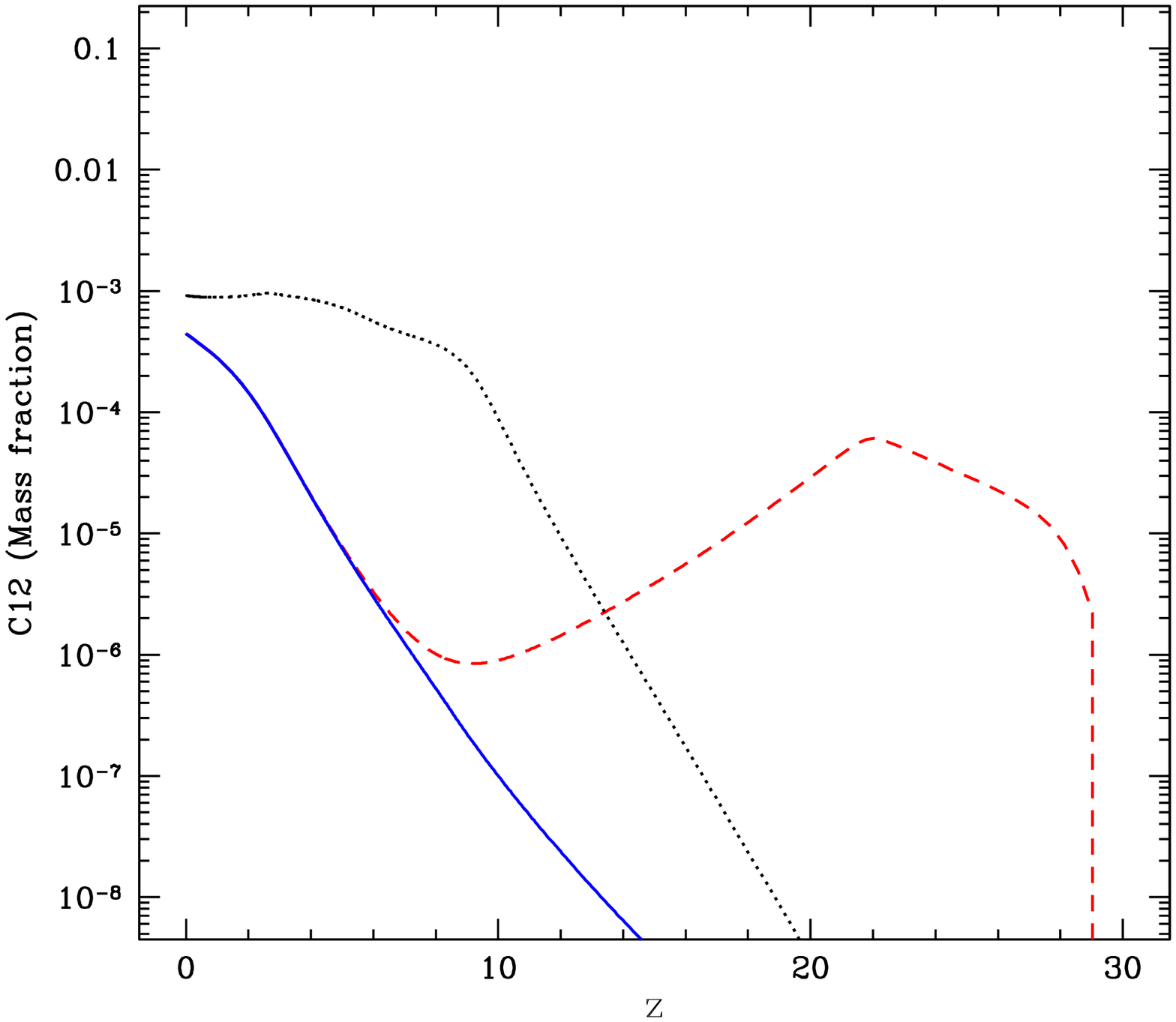, height=3.0in}&
\epsfig{file=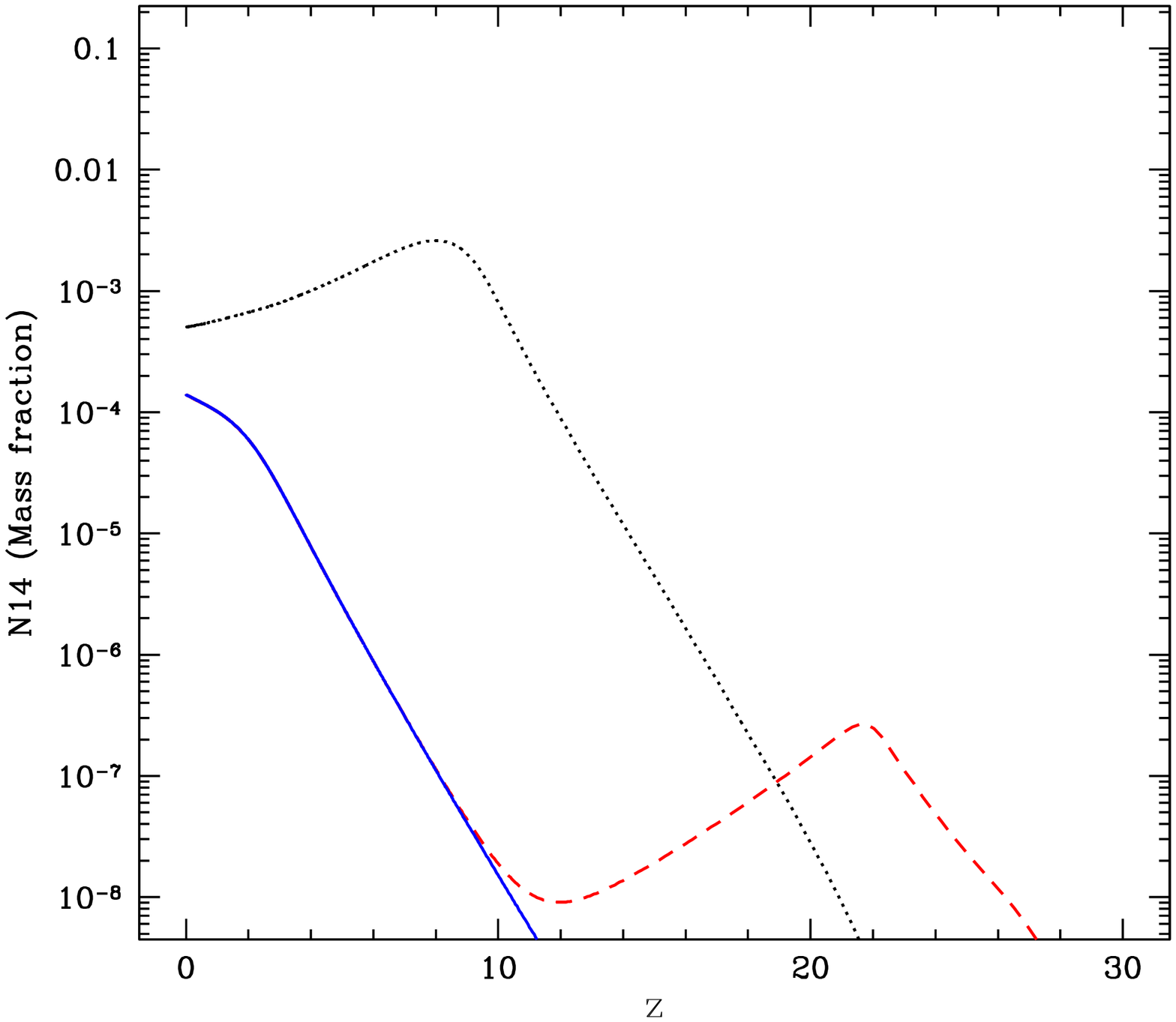, height=3.0in}\\
\end{tabular}
\end{center}
\caption{As in Fig. \ref{fig:HEOz} for Carbon (\textit{Left}), and Nitrogen (\textit{Right}).
 \label{fig:cn}
}
\end{figure}

Consider now the abundance evolution of CNOMg together with their abundance ratios.
Figures \ref{fig:cn} and  \ref{fig:omg} display the abundance evolution with $z$.
As in all of the previous figures, 
each of the following figures shows three curves corresponding to 
a model with no Pop III contribution (solid blue curve), a model with a massive Pop III component (red dashed curve) and a model with an IM Pop III component (black dotted curve). The evolutionary  behaviour of these model choices is well understood. Normal mode stars produce a moderate abundance of heavy elements at high redshift from the IMF-suppressed massive stars in that mode. The same stars, nevertheless are effective at low redshift and produce the bulk of metals observed in the Milky Way. The Massive Pop III component produces C, O and Mg at high redshift, though these abundances are diluted at low redshift by infall as structures continue to grow. The IM stellar component produces a rather specific nucleosynthetic signature: essentially C and N.

\begin{figure}[htb!]
\begin{center}
\begin{tabular}{cc}
\epsfig{file=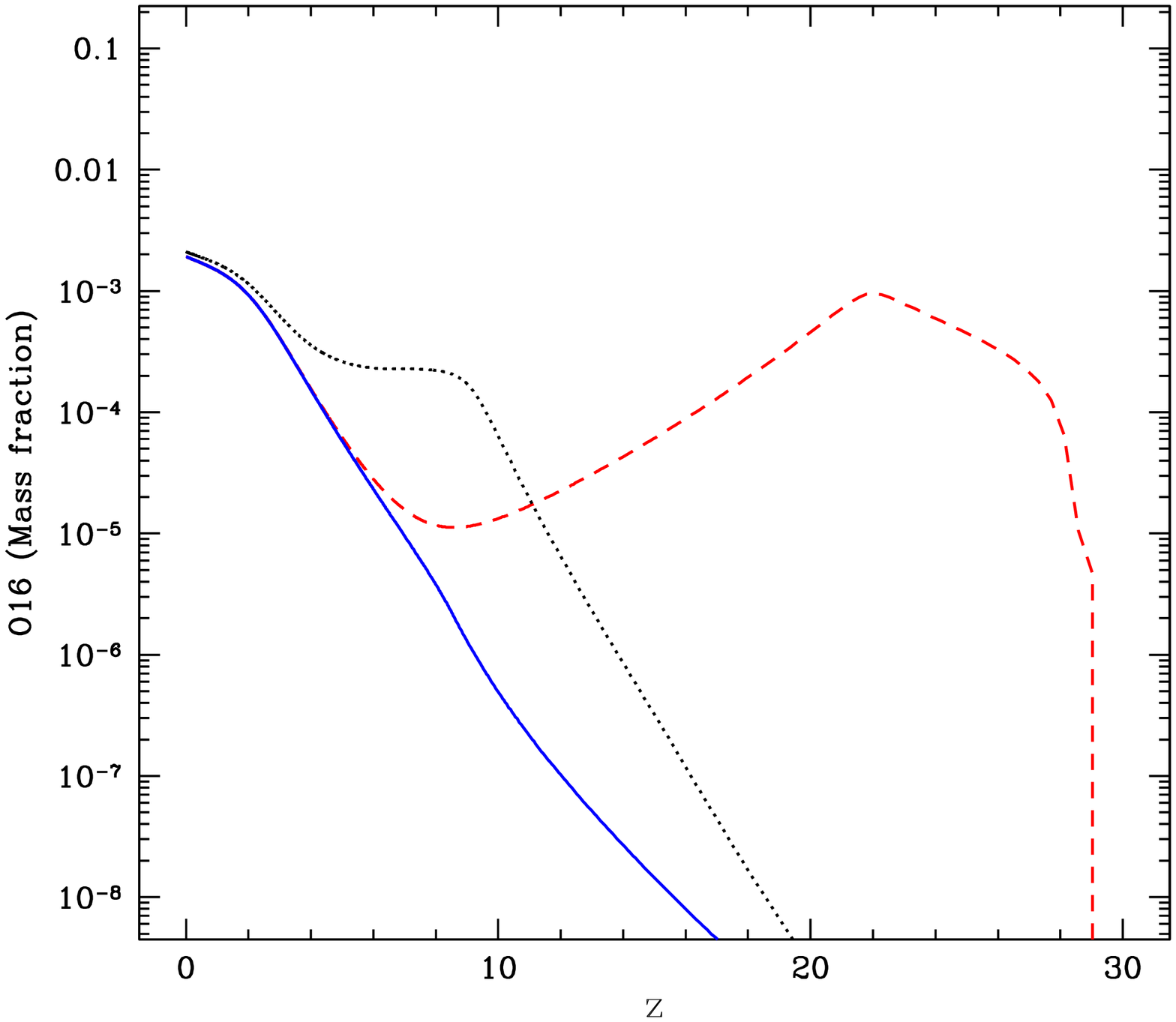, height=3.0in}&
\epsfig{file=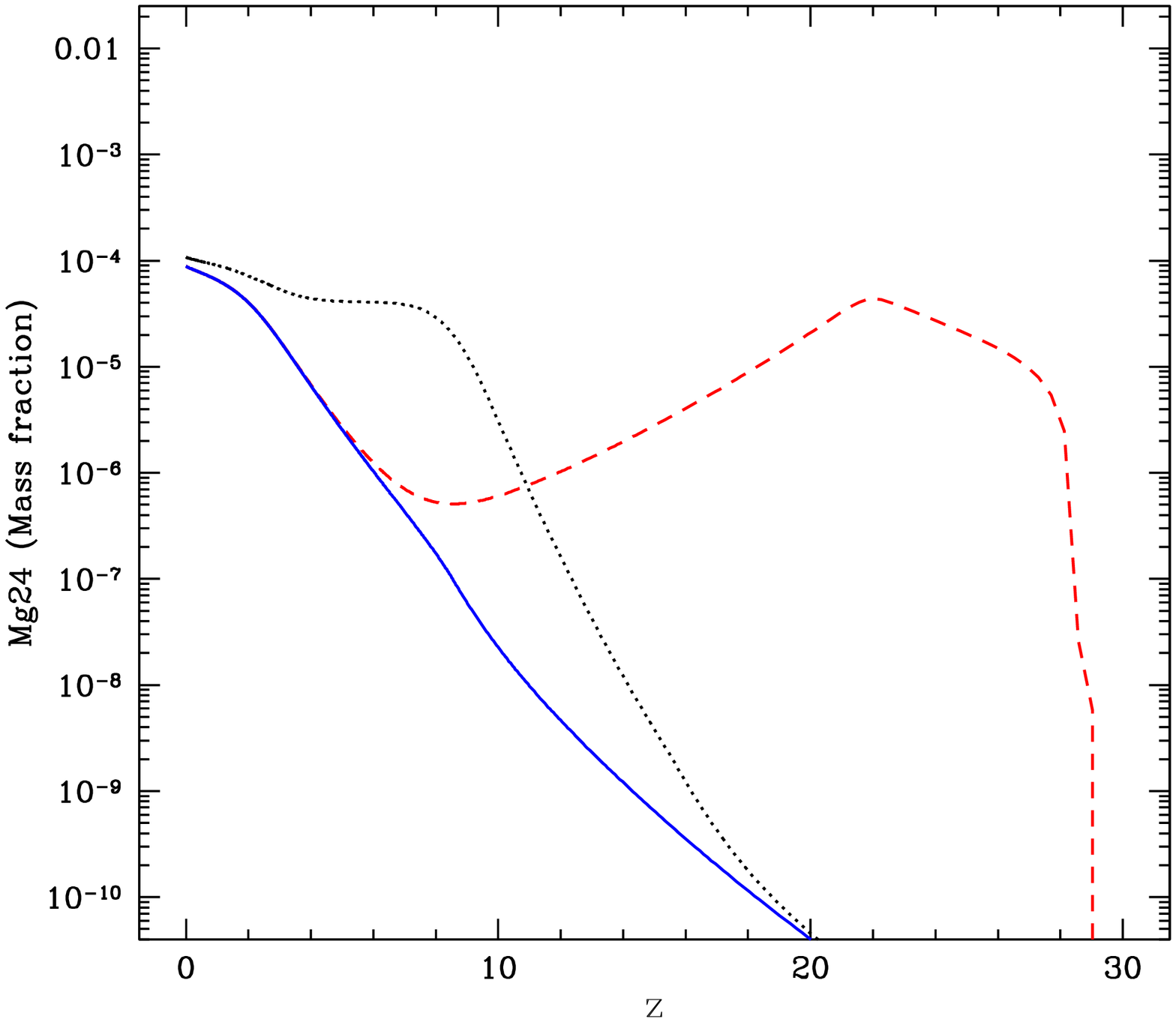, height=3.0in}\\
\end{tabular}
\end{center}
\caption{As in Fig. \ref{fig:HEOz} for Oxygen (\textit{Left}) and Magnesium (\textit{Right}). 
 \label{fig:omg}
}
\end{figure}

\begin{figure}[htb!]
\begin{center}
\begin{tabular}{cc}
\epsfig{file=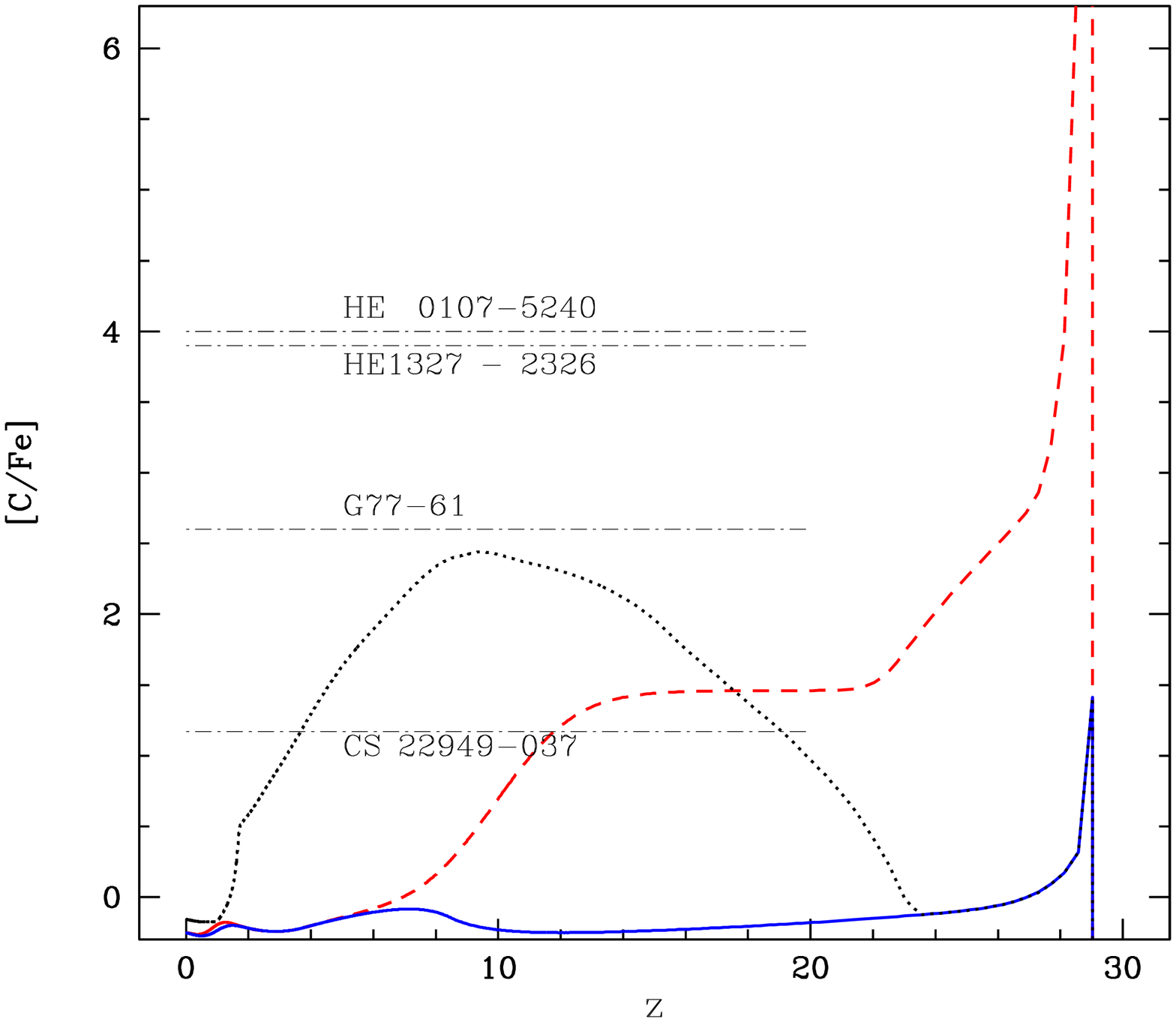, height=3.0in}&
\epsfig{file=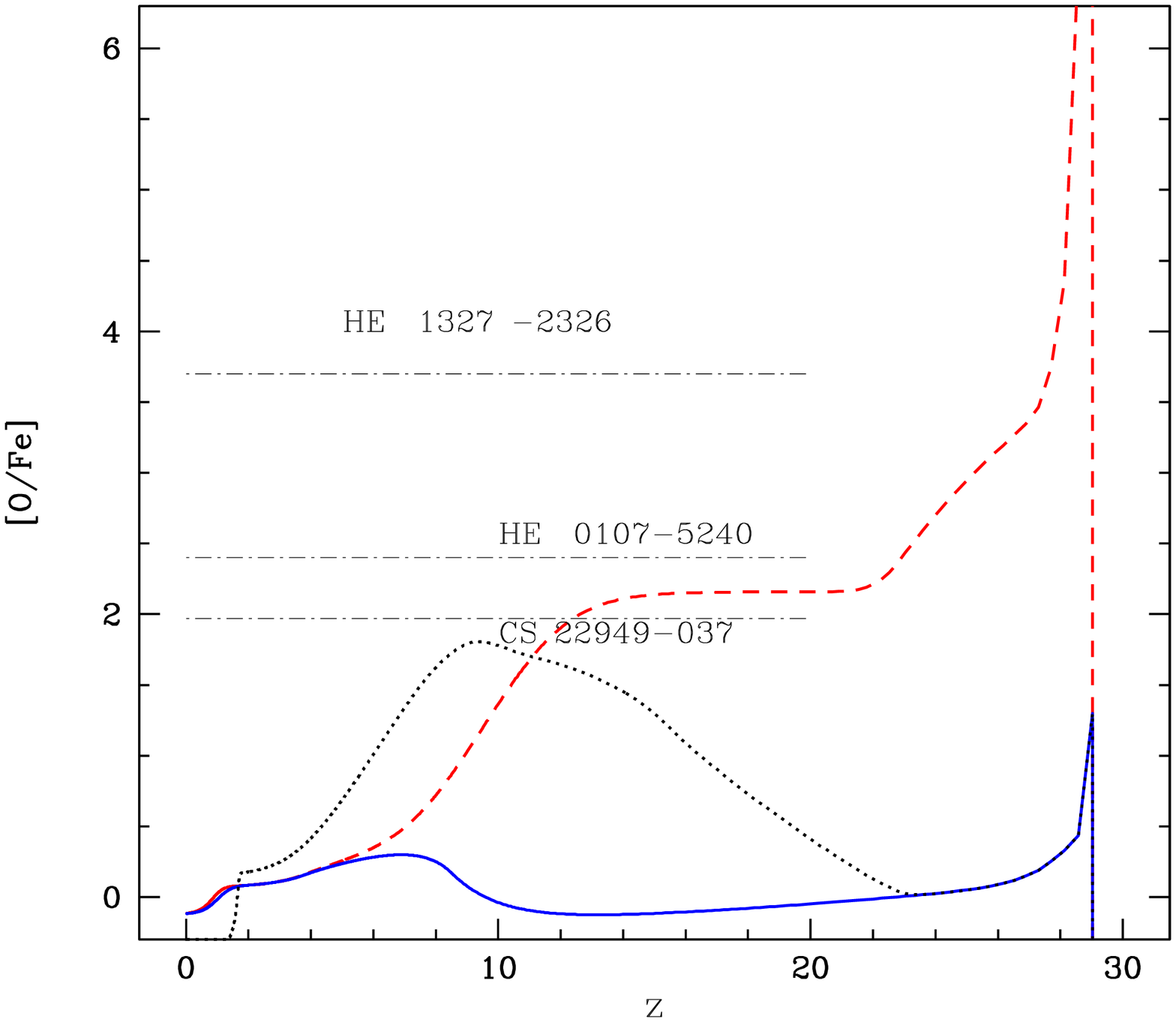, height=3.0in}\\
\end{tabular}
\end{center}
\caption{\textit{Left:} As in Fig. \ref{fig:HEOz}, the  evolution of the [C/Fe] ratio 
(relative to the solar value) as a function of redshift.
\textit{Right:} Idem for [O/Fe] ratio.
Observational data (horizontal dashed lines) represent measured abundances in the following very iron poor halo stars: CS 22949-037 \citep{depagne02}, He 0107-2240 \citep{bessel04}, He 1327-2326 \citep{frebel05} and G77-61 \citep{plez05}.
\label{fig:cofe}
}
\end{figure}

\begin{figure}[htb!]
\begin{center}
\begin{tabular}{cc}
\epsfig{file=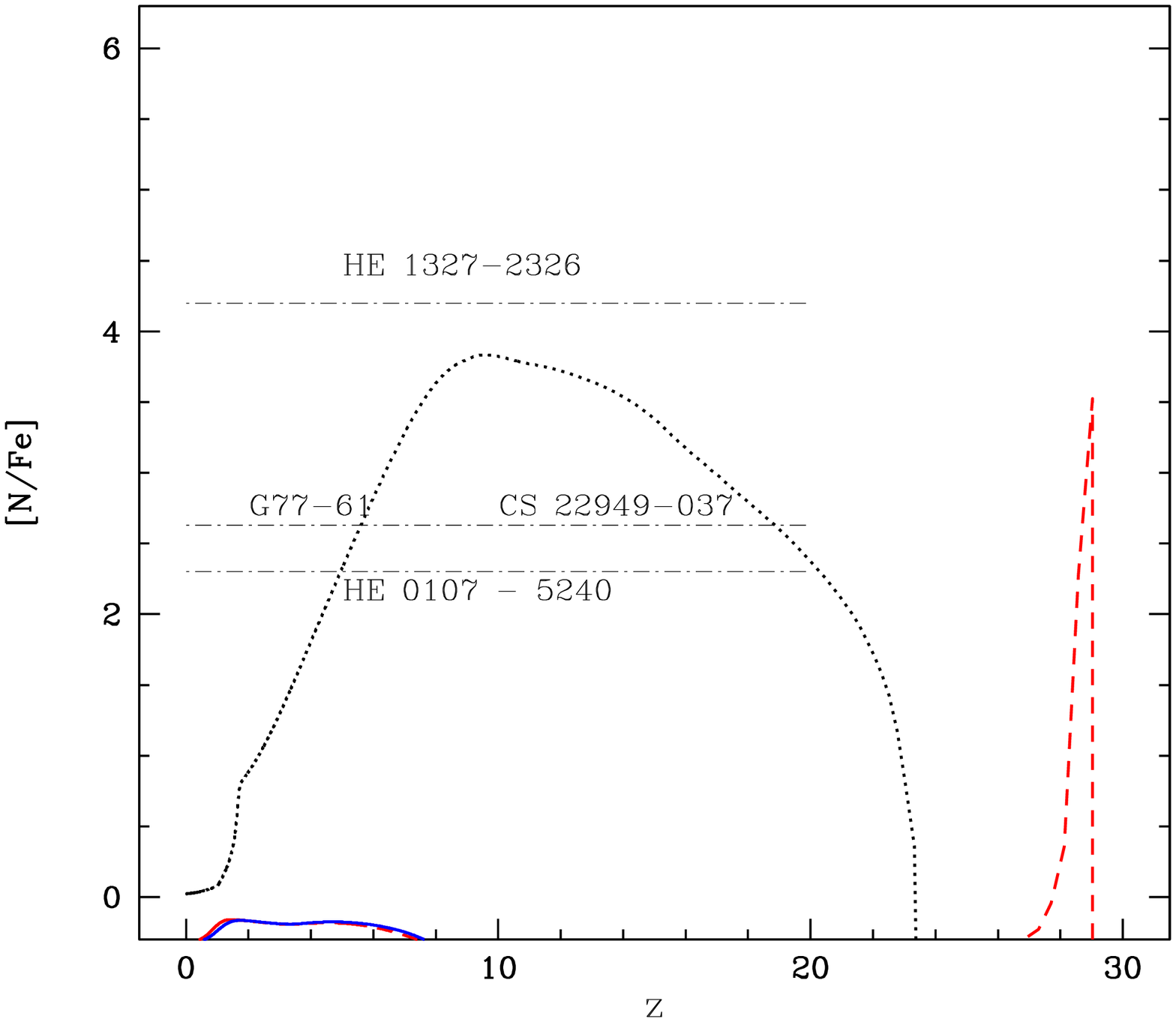, height=3.0in}&
\epsfig{file=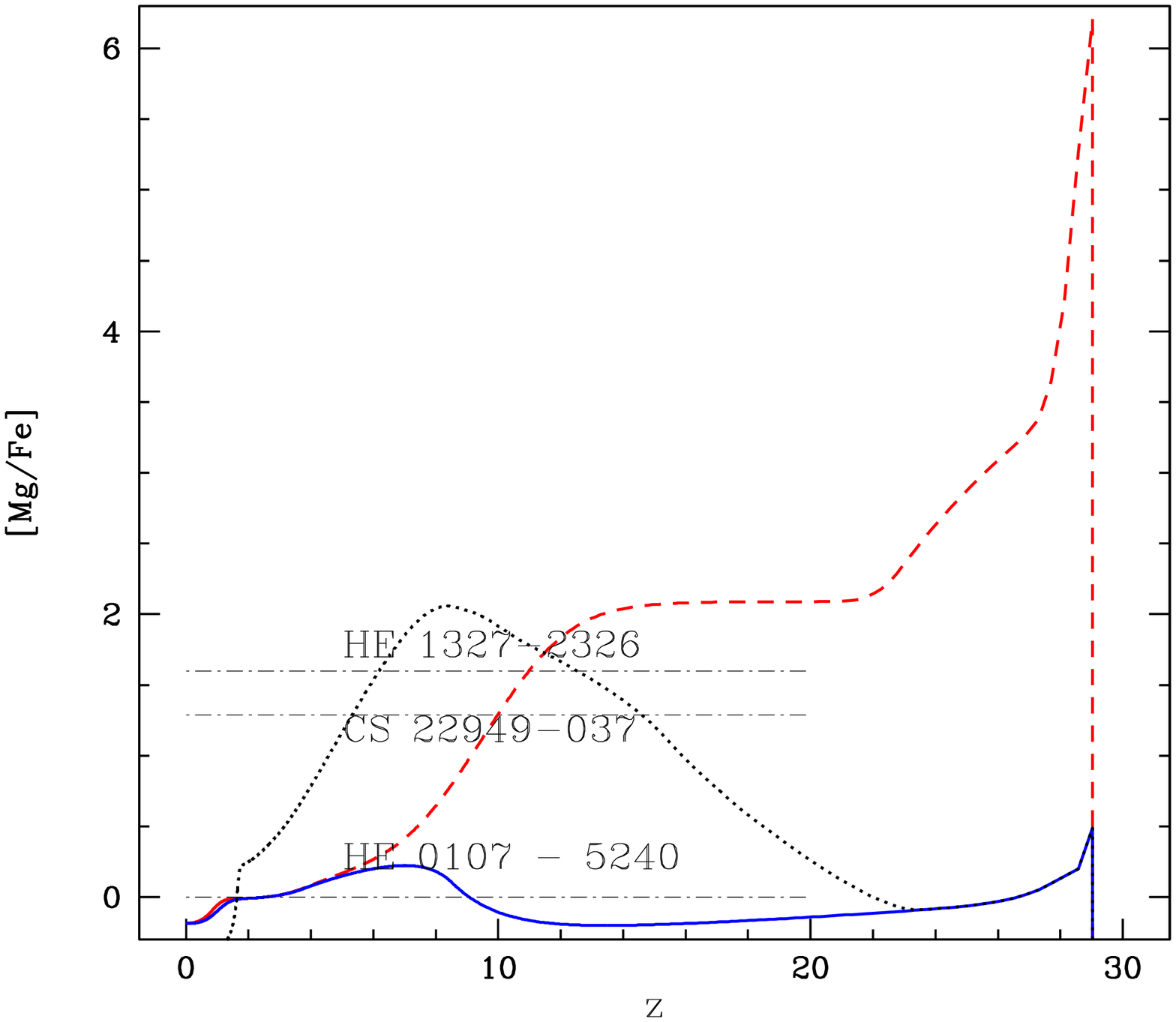, height=3.0in}\\
\end{tabular}
\end{center}
\caption{As in Fig. \ref{fig:cofe} for the [N/Fe] and  [Mg/Fe] ratios.
 \label{fig:nmgfe}
}
\end{figure}

Figures \ref{fig:cofe} and \ref{fig:nmgfe} show the evolution of the abundance ratios as a function of redshift.
As one might expect, at lower redshifts, the IM star Pop III component produces the 
highest [O/Fe], [C/Fe] , [N/Fe]. The evolution of  [Mg/Fe]  corresponds to ${}^{24}$Mg coming from massive stars and the highest
 value comes from the massive population III component.
 The ratio of the neutron-rich isotopes to ${}^{24}$Mg remains small in the Pop III model
 dominated by massive stars, but this ratio can become large ($\mathcal{O}$(1)) in the IM Pop III model.
 Also shown in these figures are the abundance ratios seen in several CEMPS.

\begin{figure}[htb!]
\begin{center}
\epsfig{file=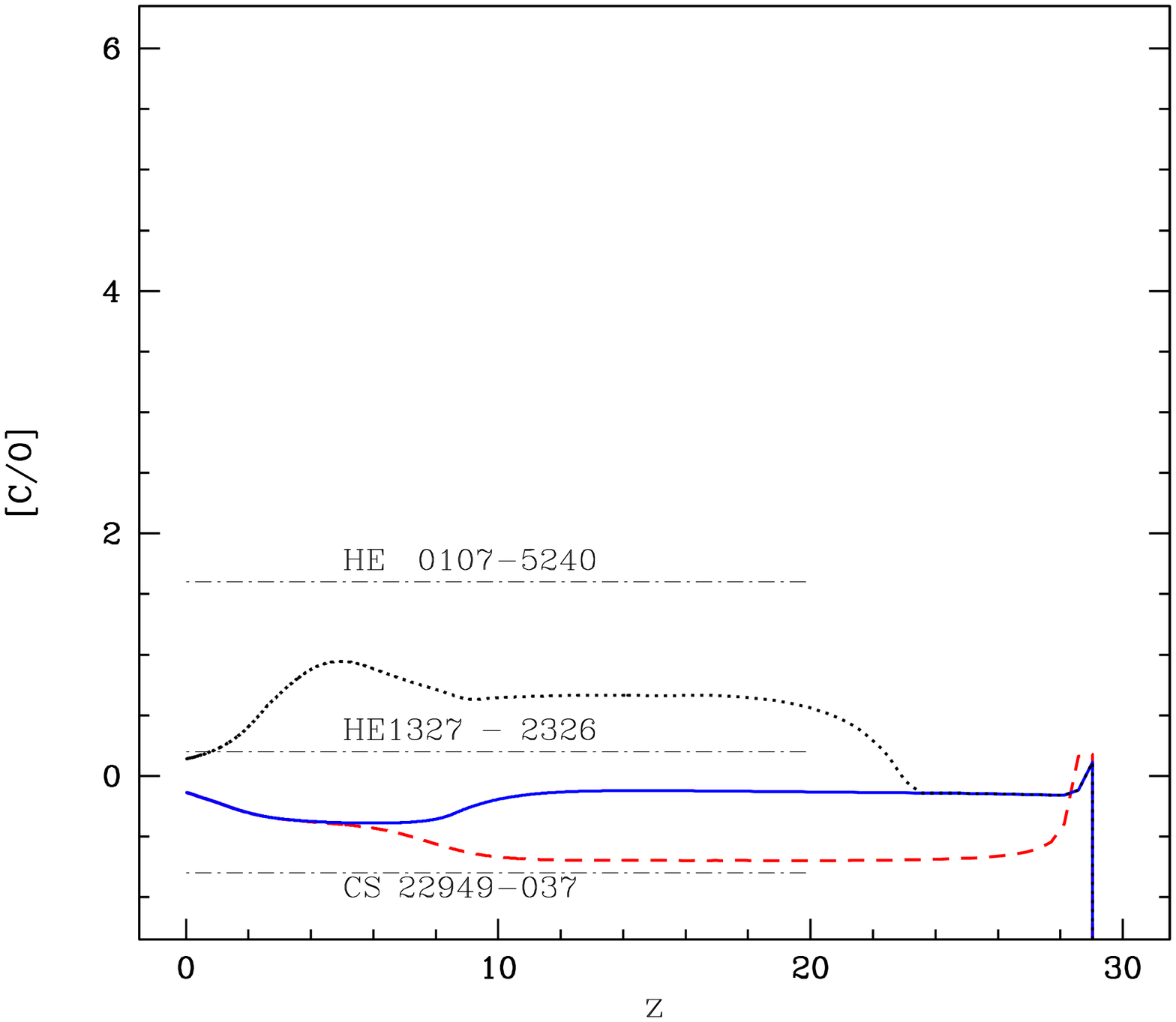, height=3.0in}
\end{center}
\caption{As in Fig. \ref{fig:cofe} for the [C/O] ratio.
 \label{fig:co}
}
\end{figure}

Surprisingly, the [C/O] ratio shown in Figure \ref{fig:co} is high in this model as oxygen is also primarily produced in more massive stars whose role is diminished in this model at intermediate redshifts.  In contrast, the massive Pop III component reaches very high ratios of [C/Fe] and [O/Fe]. Only the massive Pop III model is capable of reproducing the [O/Fe] abundance ratios observed in these stars. Similarly, the IM model has difficulty in achieving [C/Fe] ratios as high as the two HE
stars with [C/Fe] $\approx 4$ as seen in the figure, although the model does a reasonable
job of accounting for the [C/O] ratios seen in these stars.
The abundance patterns found in these CEMPS continues to confirm that the formation of massive stars at very high redshift  is plausible.

\begin{figure}[htb!]
\begin{center}
\epsfig{file=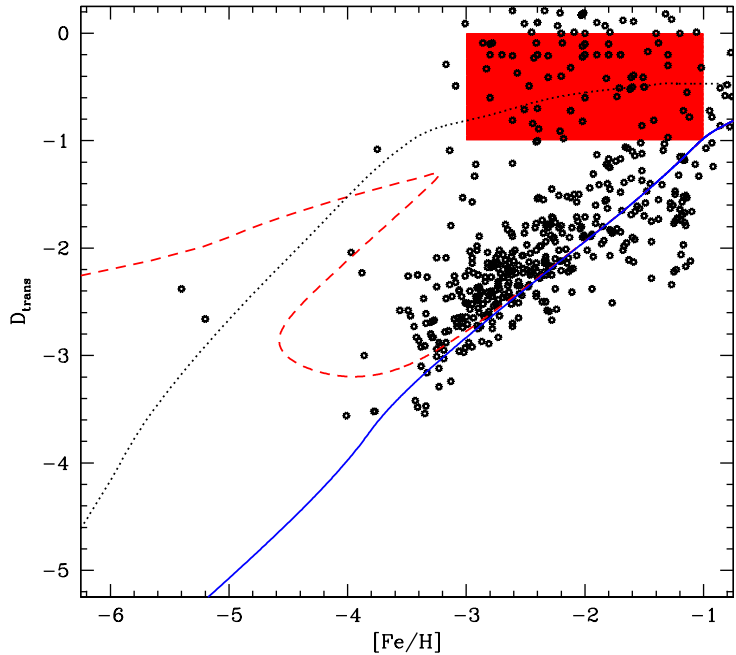, height=4.0in}
\end{center}
\caption{As in Fig. \ref{fig:HEOz}, the evolution of the D$_{\rm trans}$ parameter (transition discriminant  D$_{\rm trans}$ - see eq. (\ref{dtrans})) as a function of [Fe/H] . The black circles come from the compilation of  \cite{frebel07}. 
The red box corresponds to the region corresponding to C rich stars.
 \label{fig:dtrans}
}
\end{figure}

Finally we consider the evolution of the parameter D$_{\rm trans}$ defined above in eq. (\ref{dtrans}).
Figure \ref{fig:dtrans}  shows the D$_{\rm trans}$ parameter as a function of the metal enrichment.
Black circles come from a compilation of \citet{frebel07}. We see that this representation allows one to clearly distinguish the effects of the different choices of stellar mass ranges: the normal mode fits the bulk of standard stellar data. The massive mode can explain the few CRUMPS \citep{rollinde} and the IM stellar mode can explain the bulk of stars very enriched in C between $-3<[{\rm Fe/H}]<-1$ range ( red box). Indeed, we can see that the black dotted line (IM stars) is always much higher than  the blue one (normal mode).

In particular, for [Fe/H] $> -3$, IM stars give 
D$_{\rm trans} \ga -1$. In Figure 1 of  \citet{frebel07},
 a specific stellar population is found exactly there: C rich stars. 
It is very interesting to note that only the IM component is able to explain this part of the diagram.
Indeed, to the best of our knowledge, the yield patterns of IM stars
are the only ones which can correctly fit the abundances seen in C stars.
Thus is it tempting to interpret the metal-poor, C-rich stellar 
population as an indication of the presence of an early IM population.
One could even push this further:  of the stars in Figure \ref{fig:dtrans},
about $\sim 10 - 20 \%$
fall in the C-rich red box fit by IM stars.
To the extent that the data in the figure faithfully trace the 
ensemble of star-forming histories in the early universe, 
we would then estimate that Pop III IM stars form at $\sim 10-20\%$ of our
model rate.  Since our IM model has $\sim 10\%$ of baryons processed
through this mode, the C-rich stars would in turn imply
that $\sim 1-2\%$ of baryons participate in IM star formation.

Thus we see that C-rich halo stars seem to demand {\em some} early IM star formation.
Moreover, the large {\em scatter} in C and O abundance patterns over the
full metal-poor halo star population demand that no single nucleosynthesis (and thus
star-forming) history will suffice. Indeed, given that C-rich stars are
outliers to the main trend, the IM population appears to
be probably sub-dominant, acting only in low mass galactic structures at high z.

\subsection{Type II supernovae}

\begin{figure}[htb!]
\begin{center}
\epsfig{file=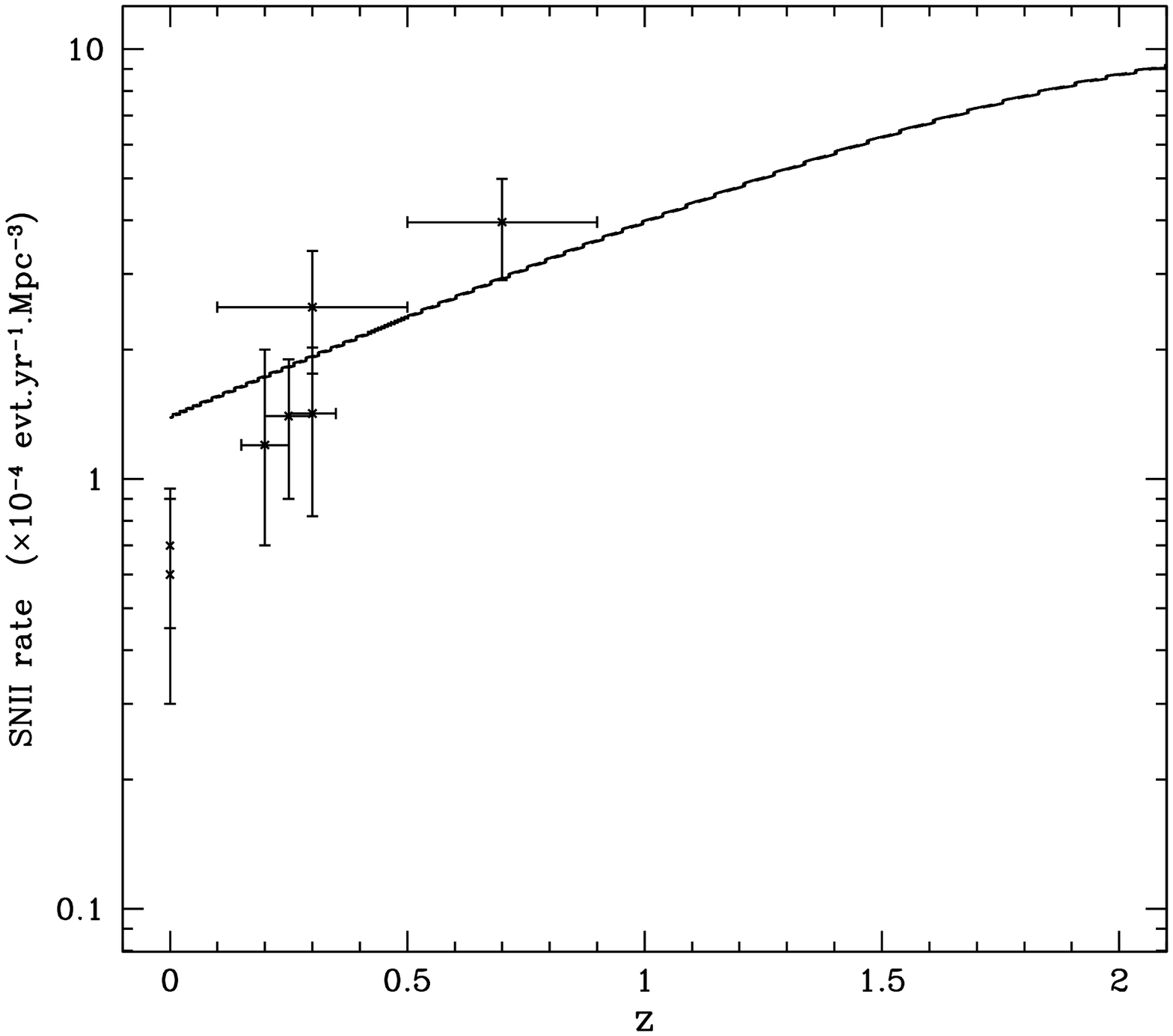, height=4.0in}
\end{center}
\caption{Evolution of the SNII rate as a function of redshift. The observed rates are taken from \citet{cap99,cap05,dal04,dal08,boti08,bazin09,li10}. The evolution is the same for all three models.
 \label{fig:SNII}
}
\end{figure}

In Figure \ref{fig:SNII},
at high redshift, 
we show the resulting evolution of the Type II SN rate along with the data at relatively low redshift.
The GOODS data for core collapse supernovae have been placed in two bins at $z = 0.4 \pm 0.2$,  and 
$z = 0.7 \pm 0.2$ (\cite{dal04,dal08}).  Their results (which have been corrected for the effects of extinction) show SNII rates which are significantly higher than the local rate 
\citep[at $z=0$;][]{cap99,li10}. 
Since the timescale between star formation and the core collapse explosion is very short, there will be
no significant contribution of Population III stars to Type II rates at any $z$.  
The SN Type II rate is directly related to the overall SFR and therefore
to the astration rate $\nu$ and the slope of the SFR at high redshift. 

The data at low $z$ is well described by the models. This result is independent of our choice of the mass range for Pop III since massive Pop III stars will only contribute to the SN rate at  very high redshift and our IM Pop III stars produce no SNII.
However, the model predicts much higher SN rates at higher redshift.  
Indeed, the SN Type II rate peaks at
a redshift $z \approx 3$ at a rate which is nearly 8 times the observed rate at low $z$. 
It will be interesting to see whether future data will be able to probe the SN rate at higher redshift.
These elevated rates may be detectable in spite of the expected increase in dust extinction due to the
early production of metals.

\subsection{Type Ia supernovae}

Our calculation of the Type Ia SN rate depends on two additional assumptions beyond
the specification of the models present in the previous section.
The Type Ia rate will depend on the fraction of low and IM mass stars which 
end up as SNIa, as well as the time delay between formation and explosion.
Furthermore, it is not clear that either of these quantities are universal constants.  
That is, they may vary with redshift, metallicity, or the size of the structure the stars are formed in.
We have assumed that the SNIa rate is proportional to the IM star SFR 
($2 - 8 \msol$). The coefficients 
adopted are $\epsilon = $ 0.02 and 0.01 for the models 2, 1 and 3 respectively.

\begin{figure}[htb]
\begin{center}
\begin{tabular}{cc}
\epsfig{file=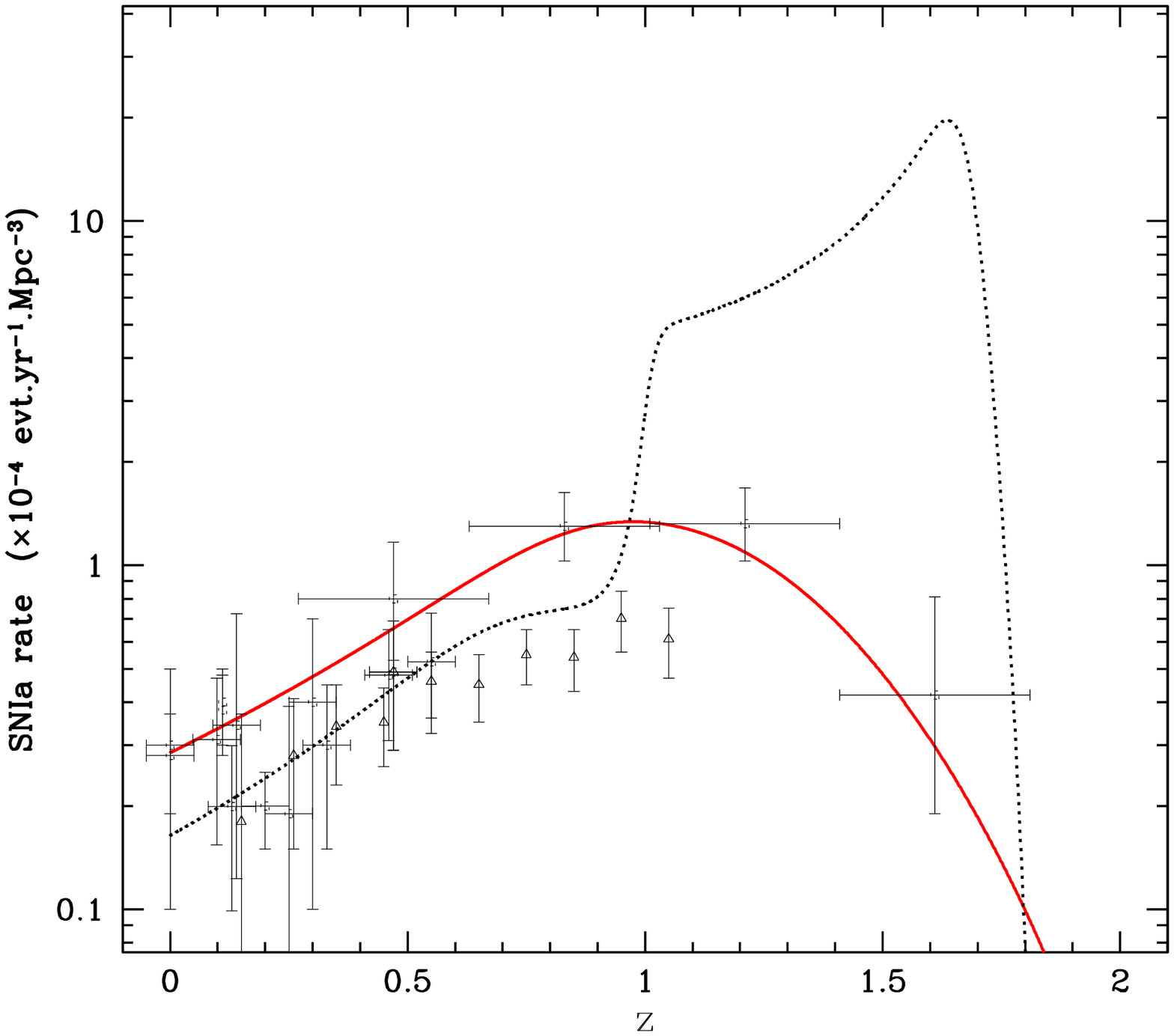, height=3.0in}&
\epsfig{file=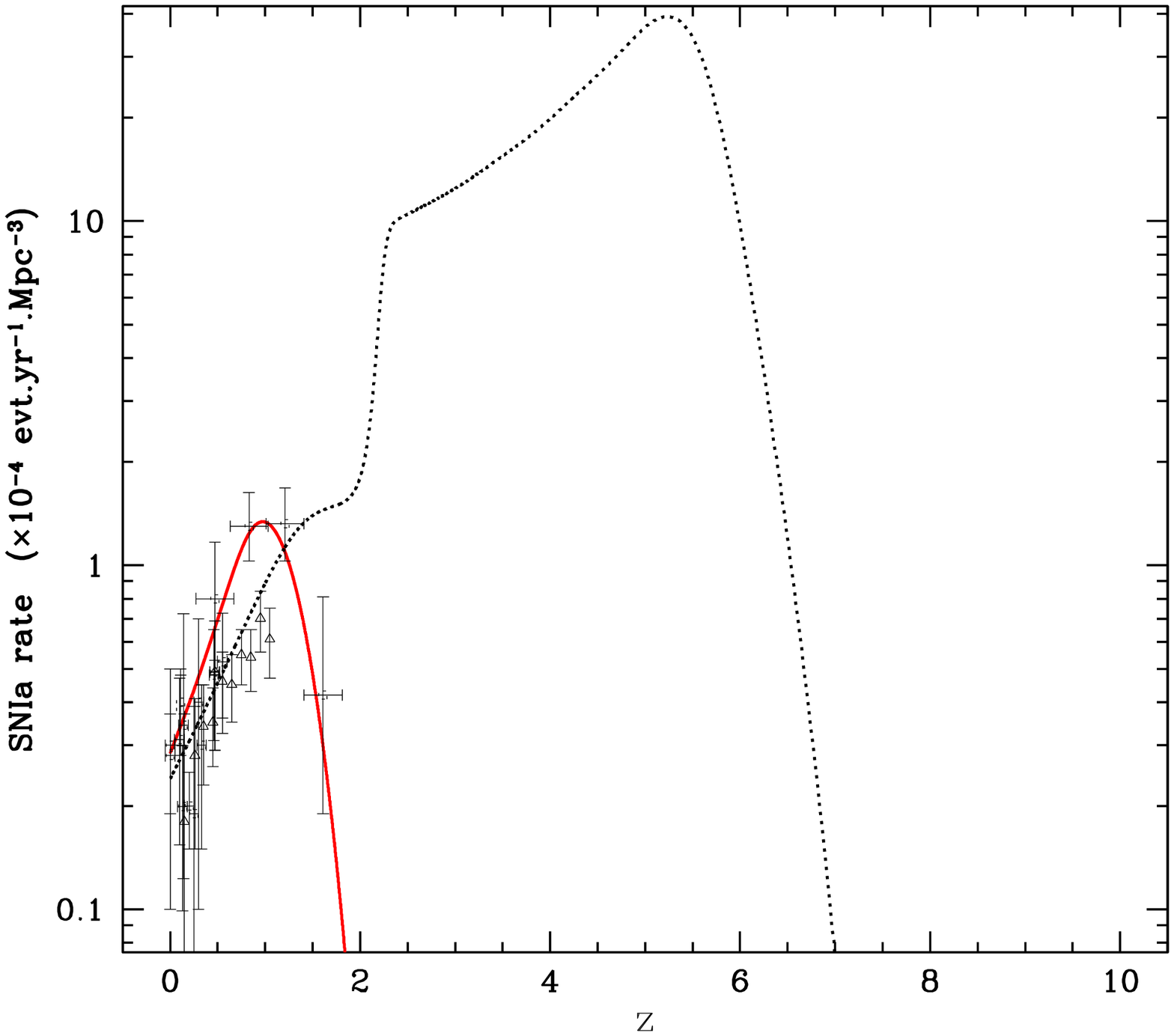,height=3.0in}\\
\end{tabular}
\end{center}
\caption{Evolution of the SNI rate as a function of redshift. The observed rates are taken from \citet{cap99,reiss00,har00,dal04,pain02,mad03,ton03,strog04,blanc04,li10,perrett10}. 
The red solid curve corresponds to the both models 1 (normal mode) and 2 (with Massive Pop III stars). The black dotted curve corresponds to model 3 (with IM stars). \textit{Left:} The assumed time delays for IM stars in model 2 and model 3 are 2.5 and 3.4  Gyr (including the lifetime of stars) respectively, and the fraction  of the white dwarfs which become SNIa are $\epsilon = $ 2 and 1 per cent, respectively. \textit{Right:} The time delay for IM stars to become SNIa in model 3  is reduced to 0.5 Gyr and $\epsilon$ is 2 per cent for both models.
 \label{fig:SNI}
}
\end{figure}

In Fig. \ref{fig:SNI}, we show the SNIa rate contrasted with the observational data.
As in the case for Type II supernovae, at high redshift, the data 
\citep[also taken from GOODS;][]{dal04,dal08} is binned
into four redshift bins at $z = 0.4$, 0.8, 1.2, and 1.6, each with a spread of $\pm 0.2$. Other data references are given in the caption. We see that, in contrast  to the case of SNII, the evolution of the SNI rate is very different for the two models with and without IM stars. 
Adding the IM stellar mode implies a high SNIa rate at high redshift.
The position of this bump depends on the time delay of the SNIa. 

Due to large uncertainties concerning the time delay, we present two cases \citep{maoz10,totani10}. We hold the delay fixed at 2.5 Gyr for normal mode stars, but 
 in the left panel of Fig. \ref{fig:SNI}, we choose a delay of 3.4 Gyr  and in the right panel we take 0.5 Gyr for model 3. While the normal mode fits the data quite well, we see that the population of IM stars
creates a significant bump in the SNIa rate at high redshift.  Unless the time delay for these
supernovae is reduced from the nominal value of 2.5 Gyr (as in the right panel of Fig. \ref{fig:SNI}) 
the SNIa rate would greatly exceed the observed rate at redshifts between 1 and 1.5.  If the delay is 
smaller, the bump in the SNIa rate is pushed to higher redshift.  Of course, it may be that for Pop III
stars, the fraction $\epsilon$ of IM stars becoming SNIa is reduced as discussed in section \ref{presn}.
 
\begin{figure}[htb!]
\begin{center}
\epsfig{file=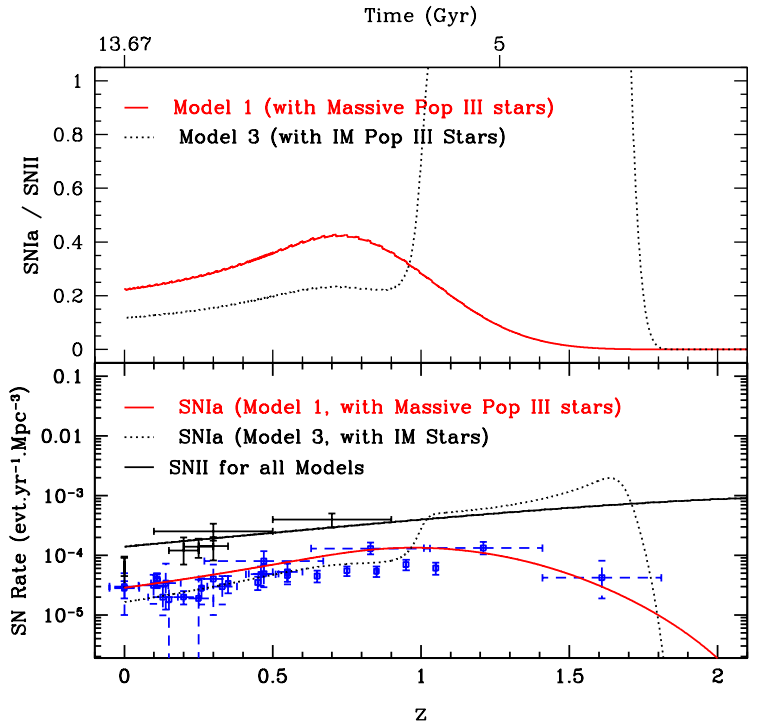, height=4.0in}
\end{center}
\caption{Evolution of the ratio of SNIa and SNII event rates as a function of redshift. (top panel) for model 1 and model 3.
In the bottom panel, the rates plotted of SNI and SNII as presented in \ref{fig:SNII} and \ref{fig:SNI}. The delay of SNIa is taken as 3.4 Gyr.
 \label{fig:Supernovae}
}
\end{figure}

In the upper panel Fig. \ref{fig:Supernovae}, the SNI/SNII ratio is plotted as a function of the redshift for models 1 and 3. In the lower panel of the figure both SNIa and SNII rates are plotted, as in \ref{fig:SNII} and \ref{fig:SNI}.
The IM stellar mode predicts that the bulk of SNIa occur at hight redshift, though this results depends on the time delay of these SN. However,  at such a high redshift  the selection efficiently drops rapidly towards zero (\cite{perrett10}) and we cannot exclude the existence of this SNIa component. In the future it would be interesting to have observational insight into low mass galaxies, which could contain sub-luminous SNIa.
An alternative, if these sub luminous SNIa are not observed we could put strong constraints on the  early helium abundance in the primitive structures.

It is interesting to note the interplay between the IM constraints from
halo star carbon abundances and from
and Type Ia SNe.  Namely, at the end of \S \ref{sect:nuke}
we argued that the existence of C-rich metal-poor halo stars
seems to {\em require} an early cosmic IM star population, but
the relative rarity of such C-rich stars also seems to imply the
IM population was sub-dominant, perhaps accounting for
$\sim 10 - 20 \%$ of early star formation.
Note that if we apply this same $\sim 10 - 20 \%$ factor to the IM predictions
for Type IA SNe in Figs.~\ref{fig:SNI} and \ref{fig:Supernovae}, 
the predictions would be brought into agreement with the observations.
Similarly, we could argue the other direction, demanding that the
IM fraction be small enough to agree with the Type Ia SN observations;
this would again give $\sim 10 - 20 \%$ of our all-IM Pop III model, and thus would 
agree with the rough statistics suggested by the counts of C-rich metal poor stars.
This rough concordance may be coincidental but is nonetheless 
both encouraging and intriguing.

\section{Discussion}
\label{end}

By definition, as the first stars formed in a metal-free environment,
Pop III stars existed and played a role in our past history.
However, the identity and mass distribution of these stars is largely unknown.
We expect that they played an important role in re-ionizing the Universe at high redshift,
as well as laid the chemical seeds for future generations of stars (population II/I).
We also know that a Salpeter IMF, typical of population I with a mass range of 
0.1 - 100 M$_\odot$ is not capable of producing a sufficient optical depth for reionization
or of producing the specific abundance patterns in EMPs. Very massive stars ($> 100$ M$_\odot$)
also fail in producing the observed abundance patterns, though they are certainly 
capable of reionizing the Universe. A top-heavy IMF for Pop III was studied in the cosmological
context in \citet{daigne2,rollinde} and while successful in many aspects, is not capable of 
producing the (albeit weak) evidence for an enhancement in \he4 at low metallicity over BBN predictions or the lagre carbon enhancements which are made manifest in the parameter
$D_{\rm  trans}$.  To the best of our knowledge, IM stars as candidates for Pop III have not been
studied in the cosmological context. 

We have shown that an early generation of IM stars is indeed capable of 
providing an effective prompt initial enrichment of He. We have also shown, perhaps surprisingly,
that these stars normalized  in abundance to the required He enrichment, are more than capable of providing a sufficient number of ionizing photons to the early
IGM.  Also as a bonus, IM stars can provide the large abundance ratios of 
C, N, O, and Mg relative to Fe in EMP stars.  However, our results indicate that 
one cannot simply assume a homogeneous and well-mixed IGM and/or ISM as
smaller structures grow to large galaxies and clusters of galaxies. The CNO abundance that
this model predicts at metallicities between $-1 <$ [Fe/H] $< -3$
far surpasses the abundances observed in the IGM.

We are thus led to an interesting dilemma which is best characterized by
the data displayed in Fig. \ref{fig:dtrans}. Recall that the discriminant $D_{\rm trans}$
represents a measure of carbon and oxygen produced in prior generations of stars.
As discussed earlier, the data show two distinct populations: the bulk of the data show
increasing  $D_{\rm trans}$ with [Fe/H] and is very well modeled by either a standard or 
top-heavy IMF; and a second population with very large  
$D_{\rm trans}$ between $-1 <$ [Fe/H] $< -3$.
The latter is well explained by our Pop III IM stellar mode.  How can one model 
explain both populations?

It is perhaps too naive to expect a homogeneous model as the type we have been describing
to explain this discrepancy.  Indeed one can imagine that the impact of the generation of IM stars
was effective only early on in the building of higher mass objects in the hierarchical 
structure formation scenario.  That is, this generation was active only in the smallest scale structures, 
here taken to be typically $10^7$ M$_\odot$. These structures and the low mass stars and remnants left behind were in some 
cases incorporated into the halos of larger objects to become galaxies. Others remain
as low mass dwarfs.  In this way, one can perhaps explain the bulk of the observations
of low and increasing $D_{\rm trans}$ in stars formed in larger scale 
objects involving a larger baryon fraction, while those stars with large $D_{\rm trans}$
would be explained by the impact of IM stars in the smallest structures formed.
Indeed, using both carbon data on metal-poor stars, and Type Ia SN observations,
we roughly estimate that a Pop III IM mode operates at
$\sim 10-20\%$ of our model rate.  Since our IM mode processes
about 10\% of baryons, this in turn implies that
$\sim 1-2\%$ of baryons participate in IM star formation.

In this interpretation, the prompt initial enrichment of \he4, which is observed largely in
dwarf irregular galaxies, was also impacted by IM stars as these objects were
evidently not incorporated into larger scale structures.  
In the context of the light element abundances, it is interesting to note further
that some of the stars which show large carbon enhancements also show deficiencies in \li7.
Furthermore, there is increasing evidence \citep{sbordone,mel} for a breakdown of the Spite plateau
at metallicities below [Fe/H] $< -3$. In \citet{sbordone}, it is argued that
the depletion (which shows considerable dispersion {\em below} the plateau)
at low metallicity may have been due to stellar astration.  The model discussed
here (given the above interpretation) can account for most of the astration
seen in these stars.  The model cannot, however, account for the discrepancy 
between the BBN prediction at the WMAP baryon density and the Spite plateau value
as speculated in \citet{piau}. Note that the depletion in D/H is probably not an issue,
as the absorption systems with measured D/H are presumably larger scale structures
for which our abundance patterns would not be expected to apply.

Unfortunately, neither of the two Pop III models delivers 
entirely satisfactory results.  While the massive mode, is capable of 
producing sufficiently high ratios of [C,O,Mg/Fe] to explain the observations 
of extremely iron-poor stars, the high ratios occur almost instantly after the first stars are born,
and thus require these stars to be born at that time at very high redshift.
This model also has difficulty in explaining the high [N/Fe] or [C/O] ratios seen.
As a result, the massive Pop III model cannot explain the high values of $D_{\rm trans}$
seen in some stars.
The model cannot account for the effective prompt initial enrichment of \he4
nor can it be used to account for the astration of \li7 below the Spite plateau.
On the other hand, the IM model for Pop III underproduces [O/Fe]
and because of the large amounts of C and N produced cannot explain the bulk of 
the values of $D_{\rm trans}$ observed, but the top panel data is explained due to the overabundances of C,N. 
 Indeed, the overproduction of 
these elements would preclude the homogeneous and
well-mixed treatment of gas in ever increasingly large scale structures.

Another distinctive feature of the IM Pop III model is the prediction
for the rate of Type Ia supernovae at high redshift.  This model naturally predicts
a large increase in the Type Ia rate {\em if} the efficiency for supernovae is constant and
the time delay is not small. 
It is relevant to speculate that a  small high mass tail in the IM population could lead to possible GRBs in the redshift range 10-20.
This would fill the ``GRB desert" between Pop II and the usual high redshift Pop III stars.

\acknowledgements
The work of KAO was supported in part by DOE grant
DE--FG02--94ER--40823 at the University of Minnesota. Plus PICS CNRS/USA.
We are indebted to F. Daigne for his permanent help.
Thanks very much to Anna Frebel for the use of her data table compilation. 
Thanks also to Yannick Mellier and 
Reynald Pain for their precious help regarding the SNIa rate observations.

\end{document}